\documentclass[pdflatex,sn-mathphys-num]{sn-jnl} 
\usepackage[utf8]{inputenc}
\usepackage[english]{babel}
\usepackage[T1]{fontenc}
\usepackage{lipsum}
\usepackage{natbib}
\usepackage{hyperref}
\usepackage{braket}
\usepackage{tocloft}
\usepackage{floatrow}
\usepackage{tikz}
\usepackage{tabularx}
\usepackage{adjustbox}
\usepackage{braket}
\usepackage{setspace}
\usepackage{qcircuit}
\usepackage{mathtools}
\usepackage{caption}
\usepackage{subcaption}
\usepackage{mathrsfs}
\usepackage{wrapfig}
\usepackage{titlesec}
\usepackage[acronym]{glossaries}
\usepackage{geometry}
\usepackage{graphicx}
\usepackage{amsmath,amssymb,amsfonts}
\usepackage{amsthm}
\usepackage{mathrsfs}
\usepackage[title]{appendix}
\usepackage{xcolor}
\usepackage{textcomp}
\usepackage{manyfoot}
\usepackage{booktabs}
\usepackage{algorithm}
\usepackage{algorithmicx}
\usepackage{algpseudocode}
\usepackage{listings}

\theoremstyle{thmstyleone}
\newtheorem{theorem}{Theorem}

\theoremstyle{thmstyletwo}

\theoremstyle{thmstylethree}

\graphicspath{ {./images/} }

\begin{document}

\title{Quantum Re-Uploading for Calorimetry: Optimized Architectures with Extended Expressivity}

\author[1,3]{\fnm{Léa} \sur{Cassé}}\email{lc480@students.waikato.ac.nz}

\author[1]{\fnm{Bernhard} \sur{Pfahringer}}\email{pfahringer.bernhard@waikato.ac.nz}

\author[1,2]{\fnm{Albert} \sur{Bifet}}\email{albert.bifet@waikato.ac.nz}

\author[4]{\fnm{Frédéric} \sur{Magniette}}\email{frederic.magniette@llr.in2p3.fr}

\affil[1]{\scriptsize{\orgdiv{The Artificial Intelligence Institute}, \orgname{University of Waikato}, \orgaddress{\city{Hamilton}, \country{New Zealand}}}}

\affil[2]{\scriptsize{\orgdiv{Télécom Paris}, \orgname{Institut Polytechnique de Paris}, \orgaddress{ \city{Palaiseau}, \country{France}}}}

\affil[3]{\scriptsize{\orgdiv{École polytechnique}, \orgname{Institut Polytechnique de Paris}, \orgaddress{ \city{Palaiseau}, \country{France}}}}

\affil[4]{\scriptsize{\orgdiv{Laboratoire Leprince-Ringuet}, \orgname{Institut Polytechnique de Paris}, \orgaddress{ \city{Palaiseau}, \country{France}}}}

\normalsize

\abstract{Near-term quantum machine learning must balance expressivity, optimization, and hardware constraints. We study quantum re-uploading units (QRUs) as compact circuits and compare them, at matched parameter count, to a standard mono-encoded variational quantum circuit (VQC) baseline. On a three-feature calorimetry classification task, we train a single-qubit QRU that outputs a scalar in $[-1,1]$ and map it to three classes via fixed thresholds. In this setting, QRUs obtain higher accuracy than the mono-encoded baseline. A controlled ablation over depth, input scaling, circuit template, optimizer, and gradient accumulation indicates that most gains occur at small depths, with diminishing returns as depth increases while training cost grows approximately linearly. To interpret these observations, we analyze reachable Fourier components and find that repeated data re-encoding expands the per-coordinate harmonic support relative to mono-encoding, consistent with a spectral activation study over random initializations. Finally, we report an end-to-end proof-of-execution of the trained model on a superconducting QPU via a cloud workflow, illustrating practical deployability under current constraints.}\keywords{Quantum machine learning, data re-uploading, Hyperparameter optimization, Particle classification, Fourier analysis, Expressivity, Calorimetry.}

\maketitle

\section{Introduction}

Quantum machine learning (QML)~\cite{Biamonte2017} has attracted substantial attention across domains, including particle physics~\cite{GNNParticleSim2023,GNSMeshNet2023}. The central idea is to leverage quantum state spaces and interference to build models that can, in principle, realize advantageous classification, regression, or clustering behaviors~\cite{zhou2021machine,steane1998quantum}. In practice, we remain in the Noisy Intermediate-Scale Quantum (NISQ) era~\cite{preskill18,bharti2022noisy}, characterized by limited qubit counts and shallow circuit depths, which constrains QML on high-dimensional, nonlinear data.

Within these constraints, the data re-uploading Unit (QRU) model~\cite{perez2019data} is an attractive option. In QRU, classical data is iteratively encoded as rotation parameters in single-qubit circuits, alternating data-encoding and trainable parametric gates, thereby improving expressivity while keeping qubit requirements minimal. A generic QRU layered ansatz can be written as:
\begin{equation}
U(\theta, x) = \prod_{j=1}^M e^{i g_j x}\, W_j\, e^{i V_j \theta_j},
\end{equation}
where $x$ is the classical input, $g_j$ and $V_j$ are traceless Hermitian generators for data-encoding and trainable gates, and $W_j$ are fixed unitaries. Unlike certain quantum neural-network ans\"atze~\cite{zhov2000quantum} and kernel-based quantum methods~\cite{schuld2021supervised}, QRU has been shown to possess a universal approximation capability under iterative data encoding, enabling rich nonlinear representations important in practice~\cite{glendinning2005bloch}.\\

We implement and assess a single-qubit QRU on a newly simulated dataset relevant to high-energy physics. This multi-class benchmark (electrons, muons, pions), to our knowledge previously unused in the literature, is aligned with ongoing efforts in QML for classification and energy prediction on collider-style data (e.g., LHC)~\cite{HGCalSimulations2021,GNNParticleSim2023}. We report classification accuracy, loss, and simulator execution time, and we conduct a systematic hyperparameter study (depth, learning rate, batch size, etc.). Given NISQ error sensitivity, we examine how these hyperparameters impact both accuracy and stability~\cite{kingma2014adam}, combining empirical tuning with preliminary global-optimization tests.\\

\noindent The contributions of this study are as follows:
\begin{itemize}
  \item We implement and evaluate a single-qubit QRU on a new, realistic simulated dataset for multi-class particle identification (electron/muon/pion), reporting accuracy, loss, and simulation execution time.
  \item We perform an extensive hyperparameter analysis of the QRU circuit and the training procedure, quantifying interdependencies (e.g., depth, learning rate, batch size) and ranking their relative impact for NISQ-suitable operation.
  \item We provide a mathematical analysis indicating that QRU attains a broader Fourier-frequency support than a widely used Variational Quantum Circuit (VQC)~\cite{cerezo2021variational}, clarifying limitations arising not only from qubit/entanglement costs but also from frequency-bound considerations~\cite{wang2021noise}.
  \item We present a parameter-matched comparison under an identical training protocol between a single-qubit QRU, a three-qubit VQC, and a compact Multilayer Perceptron (MLP), showing that QRU is competitive with MLP and consistently outperforms VQC on our task.
\end{itemize}

\noindent Section~\ref{sec:theory} presents the theoretical foundations, introducing the mathematical formulation of the QRU and comparing its expressivity to that of a VQC through Fourier analysis. Section~\ref{sec:methodology} describes the experimental methodology, including the dataset, model configuration, and computational environment used for the simulations.
Section~\ref{sec:results} reports and analyzes the results of hyperparameter optimization studies, covering both model-related and training-related parameters, followed by global optimization experiments using Bayesian and Hyperband approaches. It also includes a parameter-matched comparison between the QRU, VQC, and MLP baselines. Finally, Section~\ref{sec:conclusion} summarizes the findings and outlines perspectives for future work.

\section{Theoretical Foundations}
\label{sec:theory}

To understand why the QRU was chosen over other quantum machine learning architectures, we analyze and compare the expressive power of two quantum models trained with the same number of parameters~$P=9L$:

\begin{itemize}
    \item A \textbf{VQC} with $Q=3$ qubits and $L$ variational layers, each composed of local Euler rotations ($RX$, $RY$, $RZ$) and a ring of CNOT gates. Data are encoded only once at the beginning via single-qubit rotations $RY(x_j)$ on each qubit.
    \item A \textbf{QRU} with a single qubit ($n=1$) and $L$ repeated data–parameter blocks. Each block re-encodes the inputs through the sequence $RX(\theta_{2})\, RY(x_j)\, RZ(\theta_{0})$, applied for each input dimension $x_j$.
\end{itemize}

\noindent With $D=3$ input features and $Q=3$ qubits, both architectures have $3DL = 3QL = 9L$ trainable parameters. However, their capacity to generate distinct functional frequencies differs fundamentally. If each feature $x_j$ is re-encoded $K_j$ times through additive single-qubit rotations of the form $RY(x_j)$, then the model output admits a Fourier-type expansion \cite{schuld2021effect}:
\begin{equation}
f(\mathbf{x}) = \sum_{\mathbf{n} \in \mathcal{F}} c_{\mathbf{n}}\, e^{i(n_1 x_1 + \dots + n_d x_d)},
\end{equation}
whose Fourier support $\mathcal{F}$ satisfies $|n_j| \le K_j \quad \forall j$. Hence, the highest representable frequency in $x_j$ is bounded by the number of re-encodings $K_j$ rather than by the number of qubits. A mono-encoded VQC has $K_j=1$, leading to a limited support $\mathcal{F}_{\mathrm{VQC}} \subseteq \{-1,0,1\},$ whereas a QRU with $L$ re-upload blocks has $\mathcal{F}_{\mathrm{QRU}} \subseteq \{-L, \ldots, 0, \ldots, L\}.$ This means that the QRU can represent functions with higher-frequency components, achieving a richer functional expressivity than a mono-encoded VQC regardless of the number of qubits. \cite{xu2024frequency}\\

To illustrate this property, we performed a numerical experiment for $d=3$, $L=4$, $M=100$ random parameter initializations, and $N=96$ points along the diagonal $\mathbf{x} = (u,u,u)$ with $u \in [-1,1]$. 
We compute the discrete Fourier coefficients:
\begin{equation}
c_n^{(k)} = \frac{1}{N} \sum_{t=0}^{N-1} f(x_t;\theta^{(k)})\, e^{-i2\pi n t / N},
\end{equation}
and measure the activation rate:
\begin{equation}
p_n = \frac{1}{M} \sum_{k=1}^{M} \mathbf{1}\!\left(|c_n^{(k)}| > \varepsilon \right),
\end{equation}
with $\varepsilon = 0.05$ and the index $n$ corresponds to the DFT frequency order. Here, $p_n$ measures the fraction of models (among the $M$ realizations) that exhibit a significant spectral component at frequency $n$. It is therefore not a normalized probability distribution since a single model can activate several harmonics simultaneously, so $\sum_n p_n$ represents the average number of active frequencies per model, which can exceed one.

\begin{figure}[h!]
    \centering
    \includegraphics[width=\linewidth]{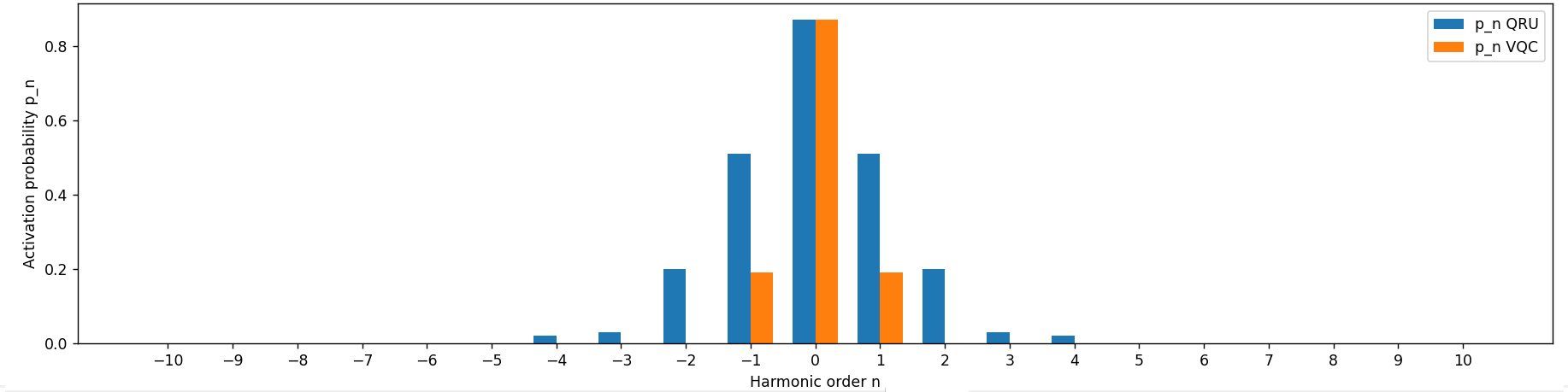}
    \caption{Empirical activation rate $p_n$ of DFT coefficients for QRU (blue) and VQC (orange) under equal parameter budgets ($P=9L$). The QRU activates harmonics up to order $\pm4$ while the VQC remains confined to $\pm1$.}
    \label{fig:spectrum-comparison}
\end{figure}

The results confirm the theoretical prediction: the QRU consistently activates all harmonics up to order $L=4$, whereas the mono-encoded VQC remains restricted to the first-order terms ($n=\pm1$). 
This agreement between theory and empirical spectra highlights the broader frequency support of the QRU, explaining its improved performance on tasks requiring fine-grained functional expressivity \cite{casse2025quantum, heimann2025learning, caro2021encoding}.

\section{Experimental Methodology}
\label{sec:methodology}

This section details the datasets, model configuration and experimental setup for evaluating QRU on a quantum simulator. We describe the data sources, baseline setups and computational environment we used in our study.

\subsection{Dataset}

The data used in this study are extracted from the OGCID/D2 simulation dataset \cite{D2dset}. It is a set of single particle interaction with a simplified highgranularity calorimeter inspired by CMS/HGCal \cite{HGCal,lobanov2020precision}.\\

This detector architecture features two main sections: the electromagnetic calorimeter (ECAL) and the hadronic calorimeter (HCAL), optimized for detecting particles of varying energies. The ECAL comprises 26 layers of lead absorbers (6.05 mm thick), while the HCAL has 24 layers of stainless steel absorbers, with 12 layers of 45 mm thickness and 12 of 80 mm, providing high resolution in capturing energy deposits. Active silicon layers are placed between absorbers, segmented into hexagonal cells with a thickness of 0.32 mm and a transverse area of approximately 1 cm2 for ECAL and 4 cm2 for HCAL \cite{GNSMeshNet2023}.\\

The D2 dataset is a collection of simulated interactions between this detector and four types of particles (muon, positive pions, electrons and photons \cite{HGCalSimulations2021, GNNParticleSim2023}) at different energies and with a flat incidence angle (aligned with the longitudinal detector axis). The dataset includes
detailed information on each particle event, namely the energy measurement in each cell but also extracted variables of interest describing the geometry and the energy repartition of the hits.\\

For this study, only one file of electrons, pions and muons has been used : the one ending with "$*\_E10-100\_theta0$". We have specifically selected some variables of interest known for their discriminating power : the total energy deposited in the ECAL, the length of the particle interaction shower and the standard deviation of the energy deposits in the HCAL.

\begin{table}[H]
\centering
\caption{Summary of the dataset used in the QRU classification experiments. All statistics are computed from the released PyTorch DataLoaders.}
\label{tab:dataset_summary}
\begin{tabular}{ll}
\hline
\textbf{Category} & \textbf{Description} \\
\hline
\multicolumn{2}{l}{\textit{Physics setup}} \\
\hline
Particle classes &
Electrons ($e$), pions ($\pi$), muons ($\mu$) \\
Energy range &
10--100 GeV \\
Incidence angle &
$\theta = 0$ (aligned with detector axis) \\
\hline
\multicolumn{2}{l}{\textit{Input features}} \\
\hline
Number of features &
3 \\
Features &
ECAL total energy; shower length; HCAL energy standard deviation \\
Input encoding &
$x \in [-\pi,\pi]^3$ \\
\hline
\multicolumn{2}{l}{\textit{Dataset size and split}} \\
\hline
Training set &
26\,334 events \\
Test set &
2\,926 events \\
Train/test split &
90\% / 10\% (event-level) \\
Class balance (train) &
$N_{-1}=9014$, $N_{0}=8289$, $N_{+1}=9031$ \\
Class balance (test) &
$N_{-1}=985$, $N_{0}=972$, $N_{+1}=969$ \\
\hline
\multicolumn{2}{l}{\textit{Implementation details}} \\
\hline
Data format &
Pre-generated PyTorch DataLoaders \\
Loader files &
\texttt{qml\_essai\_train\_loader.pth}, \texttt{qml\_essai\_test\_loader.pth} \\
\hline
\end{tabular}
\end{table}

\subsection{Baseline Setups and Computational Environment}

\hspace{1em} Before analyzing individual hyperparameters, we performed a variability study over 50 independent runs to quantify the dispersion of test accuracy and loss. This provides an estimate of the experimental uncertainty (error bars) that will contextualize subsequent hyperparameter analyses. Each run used a new random initialization and dataset reshuffling.\\

Unless otherwise specified, all experiments are based on a common baseline configuration designed for stability and reproducibility. The baseline corresponds to a single-qubit QRU with:
Huber loss, Adam optimizer, learning rate $5\times10^{-5}$, input normalization in $[-\pi,\pi]$, rotation sequence $\mathrm{RX}\!-\!\mathrm{RY}\!-\!\mathrm{RX}$, three trainable parameters per input feature, circuit depth $L=10$, and no explicit batching (full-batch updates). 
When an individual hyperparameter is analyzed, only that parameter differs from this baseline, while all others remain fixed.\\

The 3-class particle identification (electron, muon, pion) is cast as a scalar regression: each model outputs $y\in[-1,1]$, trained with a regression loss against targets $\{-1,0,+1\}$.
Class labels are recovered by thresholding at $\{-0.33,+0.33\}$.\\

Using a single expectation value to solve a three-class problem requires a fixed thresholding rule. This framing is intentionally lightweight and hardware-friendly, but it is less directly comparable to standard softmax cross-entropy baselines and can introduce sensitivity to threshold placement across random seeds or data shifts. Accordingly, throughout the paper we interpret the thresholded QRU as a minimal proof-of-concept and avoid over-claiming; extending the model to produce multiple logits (e.g., with multiple measurement heads and cross-entropy training) is a natural next step.
 We report test accuracy and final loss, together with a trainability metric defined as the area under the training-loss curve across epochs (Eq.~(\ref{eq:trainability})).\\

For the single-qubit QRU, each re-uploading layer $i=1..L$ applies, for each feature $x_j$ ($j=1..D$), the sequence 
$\mathrm{RX}(\theta^{(i)}_{j,0})$–$\mathrm{RY}(\theta^{(i)}_{j,1}x_j)$–$\mathrm{RX}(\theta^{(i)}_{j,2})$, followed by a $\langle Z\rangle$ measurement, yielding $P_{\text{QRU}}=3DL$ trainable parameters. The three-qubit VQC uses a one-shot encoding $\mathrm{RY}(x_0),\mathrm{RY}(x_1),\mathrm{RY}(x_2)$, followed by $L$ variational layers of local Euler rotations on each qubit and a ring of CNOTs; aggregation is done on $q_0$ with readout $\langle Z_0\rangle$ ($P_{\text{VQC}}=3QL$ with $Q=3$). The classical MLP is a two-layer network $3\!\rightarrow\!h\!\rightarrow\!1$ with a $\tanh$ head ($P_{\text{MLP}}=5h+1$). Parameter-matched runs use a shared depth $L$ for the quantum models ($P=9L$) and set $h$ so that $5h+1\approx9L$.\\

Unless stated otherwise, hyperparameter sweeps run for 30 epochs, while the parameter-matched QRU/VQC/MLP comparison uses Adam with a fixed learning rate $5\times10^{-5}$ for 10 epochs under identical logging. For global optimization, we apply Bayesian optimization (\texttt{skopt.gp\_minimize}, $n_\text{calls}=50$, UCB acquisition with $\kappa=4$) over depth, learning rate, loss, and optimizer, and Hyperband with fANOVA-style importance analysis and priors/surrogates when applicable.\\

All quantum models are simulated with \texttt{default.qubit}, while \texttt{PennyLane-Lightning} was tested for potential GPU acceleration. \cite{asadi2024hybrid,chen2024quantum} Small circuits achieved up to $\sim2.5\times$ speed-ups on GPU, whereas larger circuits did not significantly benefit due to shared cloud resources. For the main experiments, we executed the training scripts remotely on the LLR computing cluster, as several runs required multiple days of continuous computation to complete. \cite{llr_cluster} We record wall-clock execution times for all circuit depths to ensure consistent runtime comparisons across experiments.

\newpage
\section{Results and Hyperparameter Optimization Studies}
\label{sec:results}

This section presents the performance results of the QRU model on the particle classification task. We first analyze the impact of individual hyperparameters, organized into model-related and training-related parameters. Then, we explore the effects of global optimization techniques, including Bayesian optimization and initial tests using Hyperband and investigate correlations between key hyperparameters.

\subsection{Impact of individual hyperparameters}
This subsection examines the effect of model-related hyperparameters, specifically circuit depth, input normalization, rotation gates and the number of trainable parameters per input on the classification accuracy and computational efficiency of the QRU model.\\

In order to assess the variability of the QRU model, we analyzed the test accuracy and loss across 50 independent runs. For each run, we reshuffled the dataset and randomly initialized the parameters $\theta$ following a Gaussian distribution centered around 0.5. The plots below illustrate the test accuracy and loss for these runs:

\begin{figure}[H]
\centering
\includegraphics[width=\textwidth]{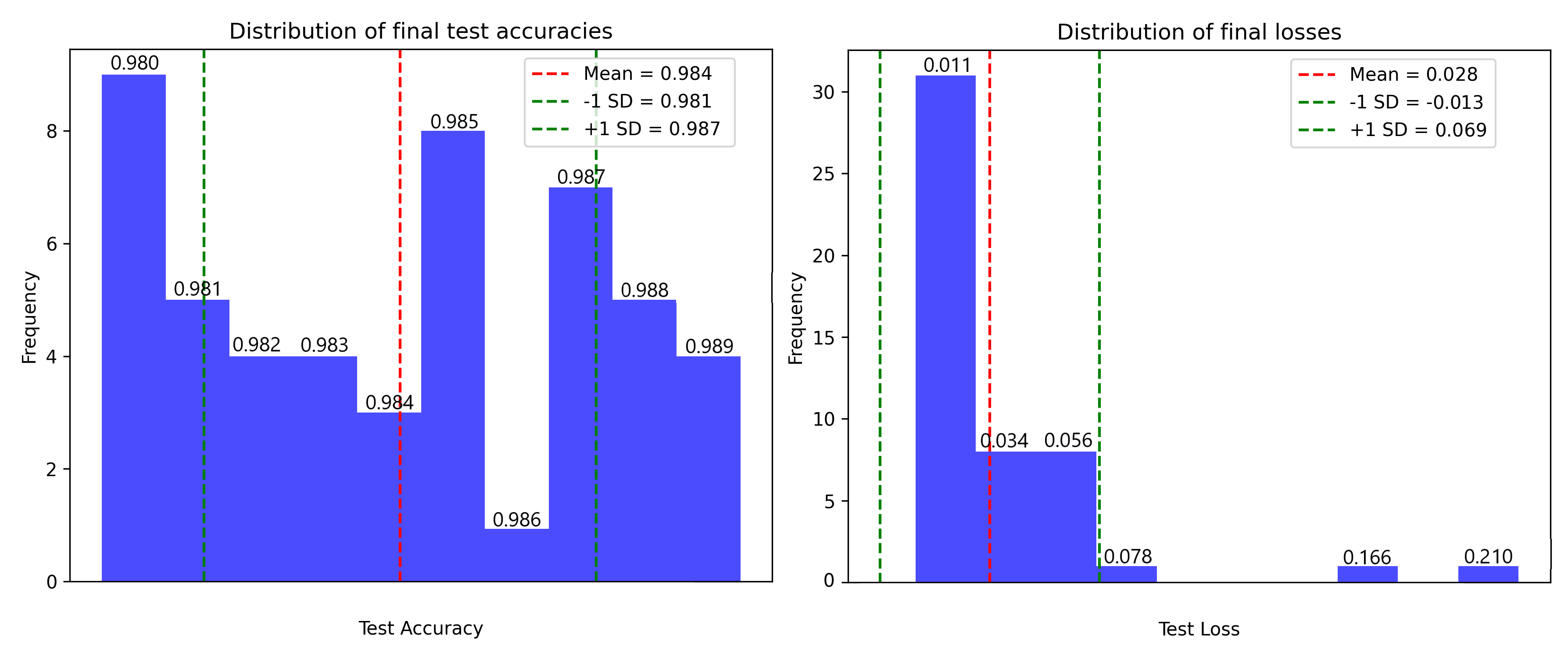}
\caption{Left: Final test accuracy; right: Final loss over 50 runs}
\label{fig:test_acc_variability}
\end{figure}

For the test accuracy, the mean is 0.98 with a variability of 0.002881, indicating that the results are highly consistent and centered around this value. On the test losses side, the mean is 0.028 with a variability of 0.041. \\

Having established the variability in the QRU's performance for classification task, we now move on to analyze how the model behaves when adjusting the depth of the quantum circuit, a critical hyperparameter for model expressivity and learning capacity.

\subsubsection{Model hyperparameters}

\hspace{1em} Having established the baseline configuration and variability range, we now examine the influence of model-specific hyperparameters, beginning with circuit depth as a primary driver of expressivity.

\paragraph{Circuit depth}\hfill

\vspace{0.5cm}

The depth of the circuit represents the number of times the data is re-uploaded into the quantum circuit. It's important to note that a depth of 1 represents a quantum circuit that has not re-uploaded the input and that means it's not technically a QRU.\\

\begin{figure}[H]
\centering
\includegraphics[width=\textwidth]{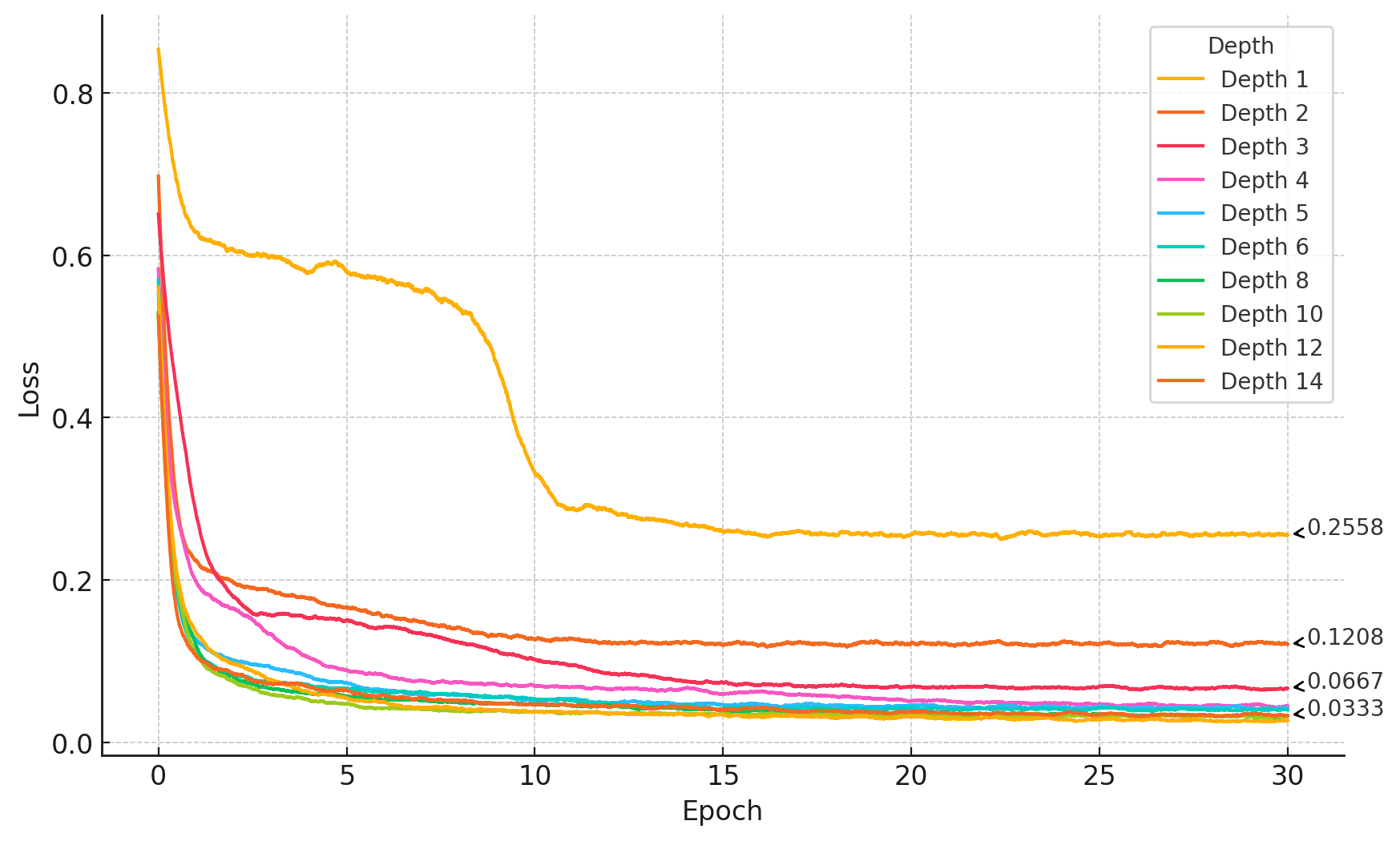}
\caption{Evolution of the loss over epochs for different circuit depths.}
\label{fig:loss_depths}
\end{figure}

Figure \ref{fig:loss_depths} shows the evolution of the loss over 30 epochs for different values of depth. We observe a significant decrease in the loss as the depth is increasing, particularly between depths 1 and 3, where the loss drops from 0.2558 to less than 0.0667. Beyond a depth of 4, the loss stabilizes around 0.03, indicating that increasing the depth beyond this threshold does not significantly improve the model's performance in terms of loss minimization.

\begin{figure}[H]
\centering
\includegraphics[width=\textwidth]{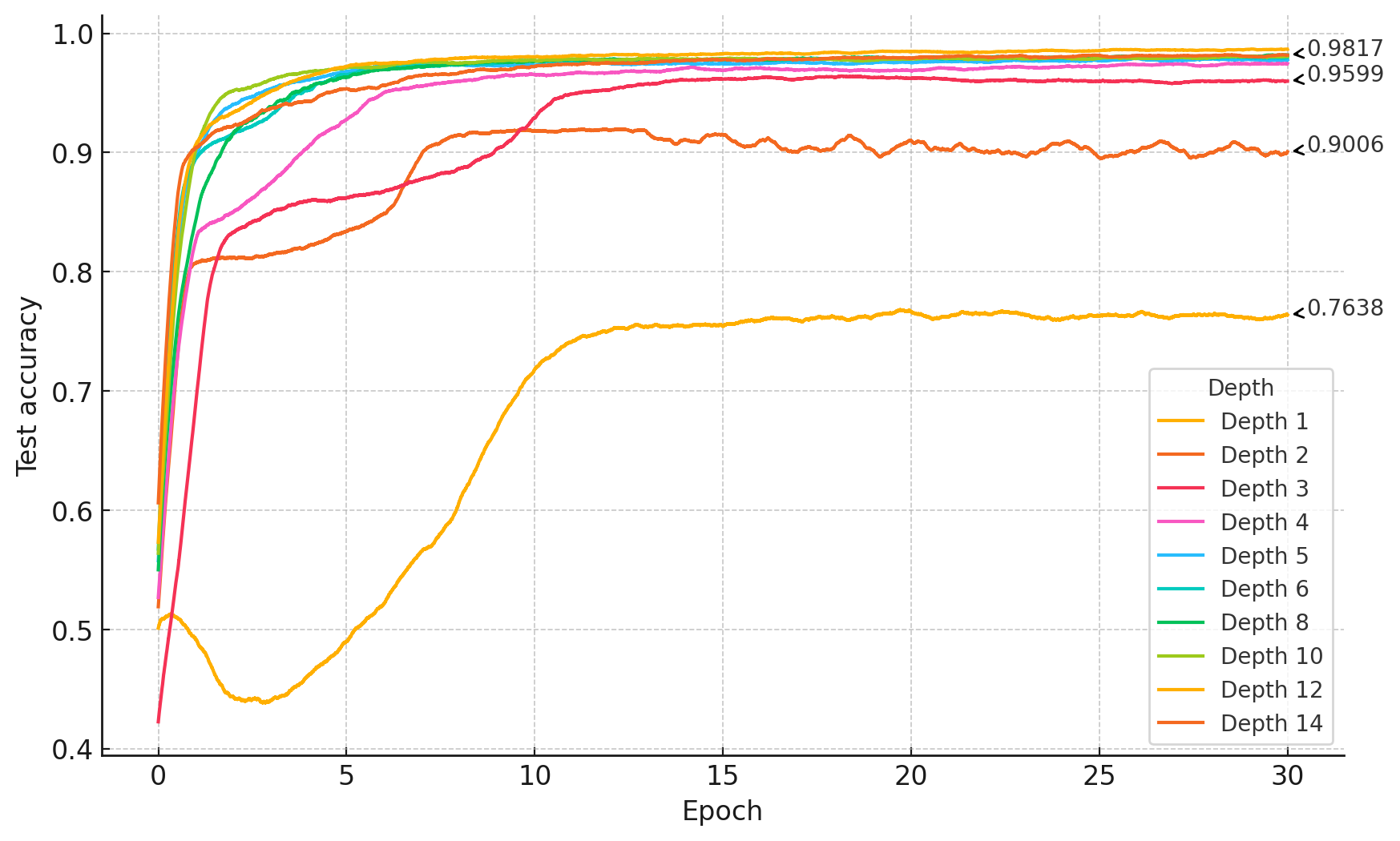}
\caption{Evolution of test accuracy over epochs for different circuit depths.}
\label{fig:test_acc_depth}
\end{figure}

Next, Figure \ref{fig:test_acc_depth} shows the evolution of test accuracy over epochs for different depth values. Similarly, we observe a rapid improvement in accuracy with shallow depths (from depth 1 to depth 3), with accuracy approaching 0.98 from a depth of 4 onwards. Again, higher depths yield only marginal gains in accuracy.

\begin{figure}[H]
\centering
\subfloat[Final loss vs depth\label{fig:final_loss_depth}]{
    \includegraphics[width=0.32\textwidth]{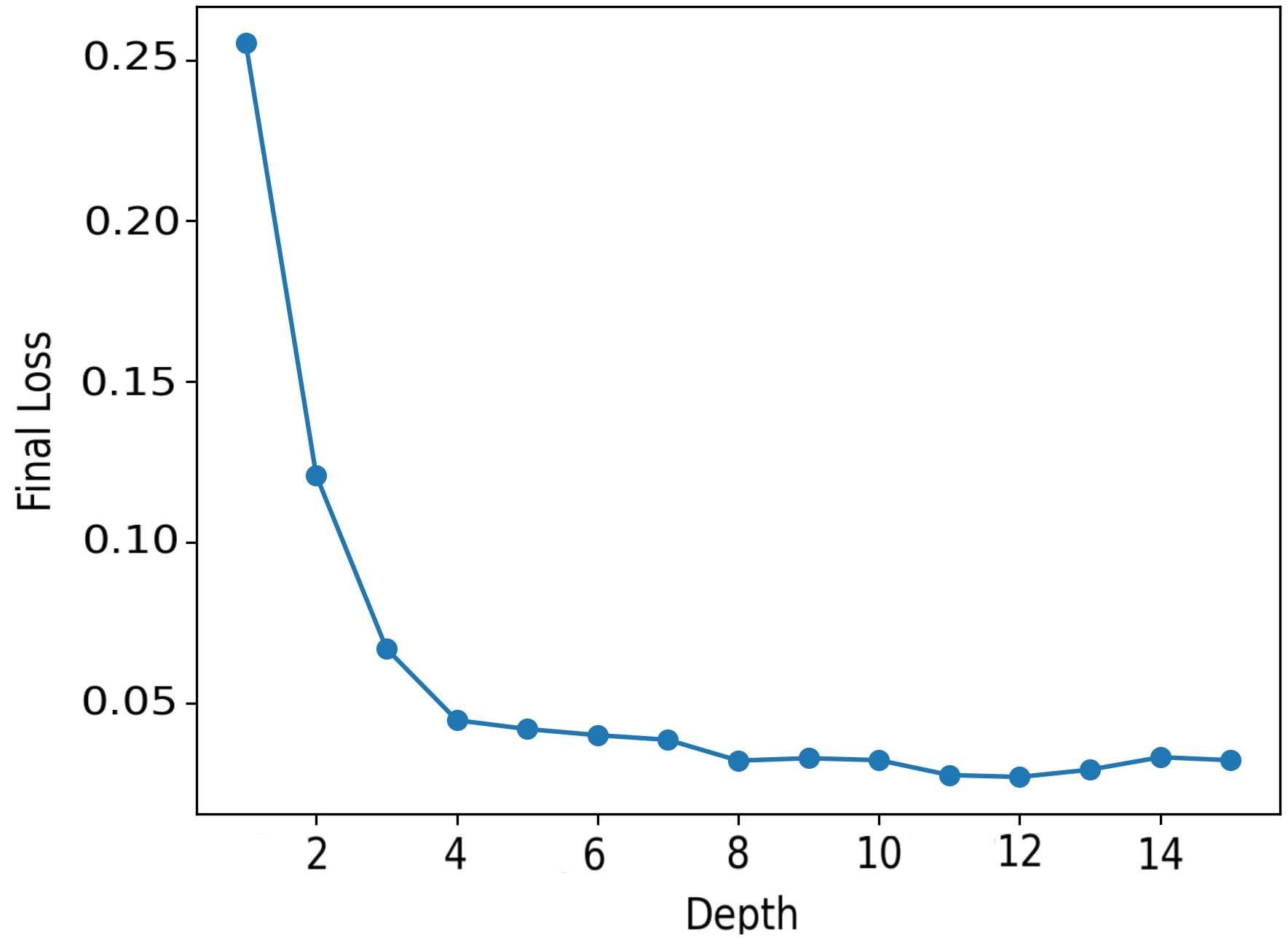}
}
\subfloat[Final test accuracy vs depth\label{fig:final_acc_depth}]{
    \includegraphics[width=0.32\textwidth]{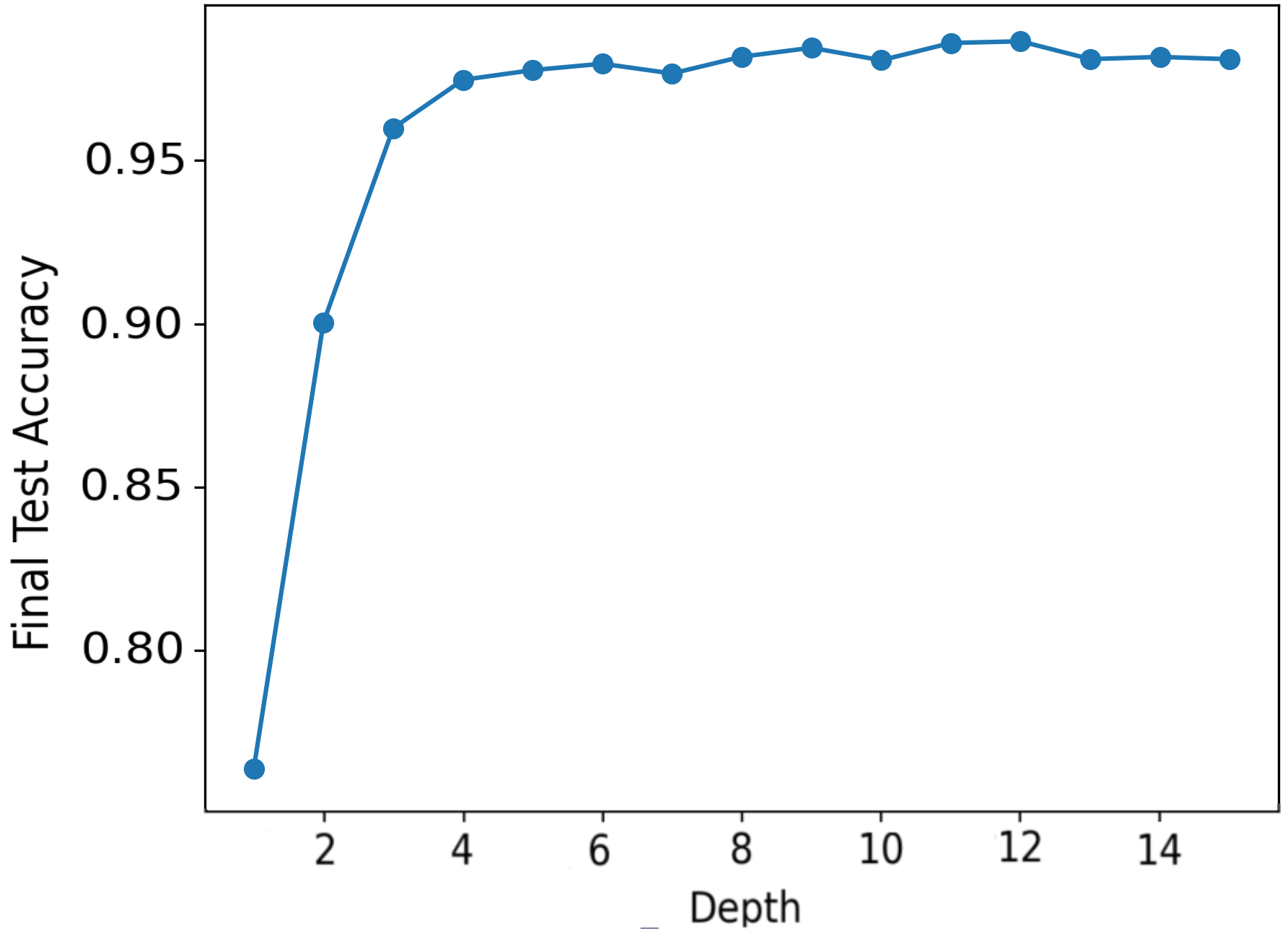}
}
\subfloat[Execution time vs depth\label{fig:execution_time_depth}]{
    \includegraphics[width=0.32\textwidth]{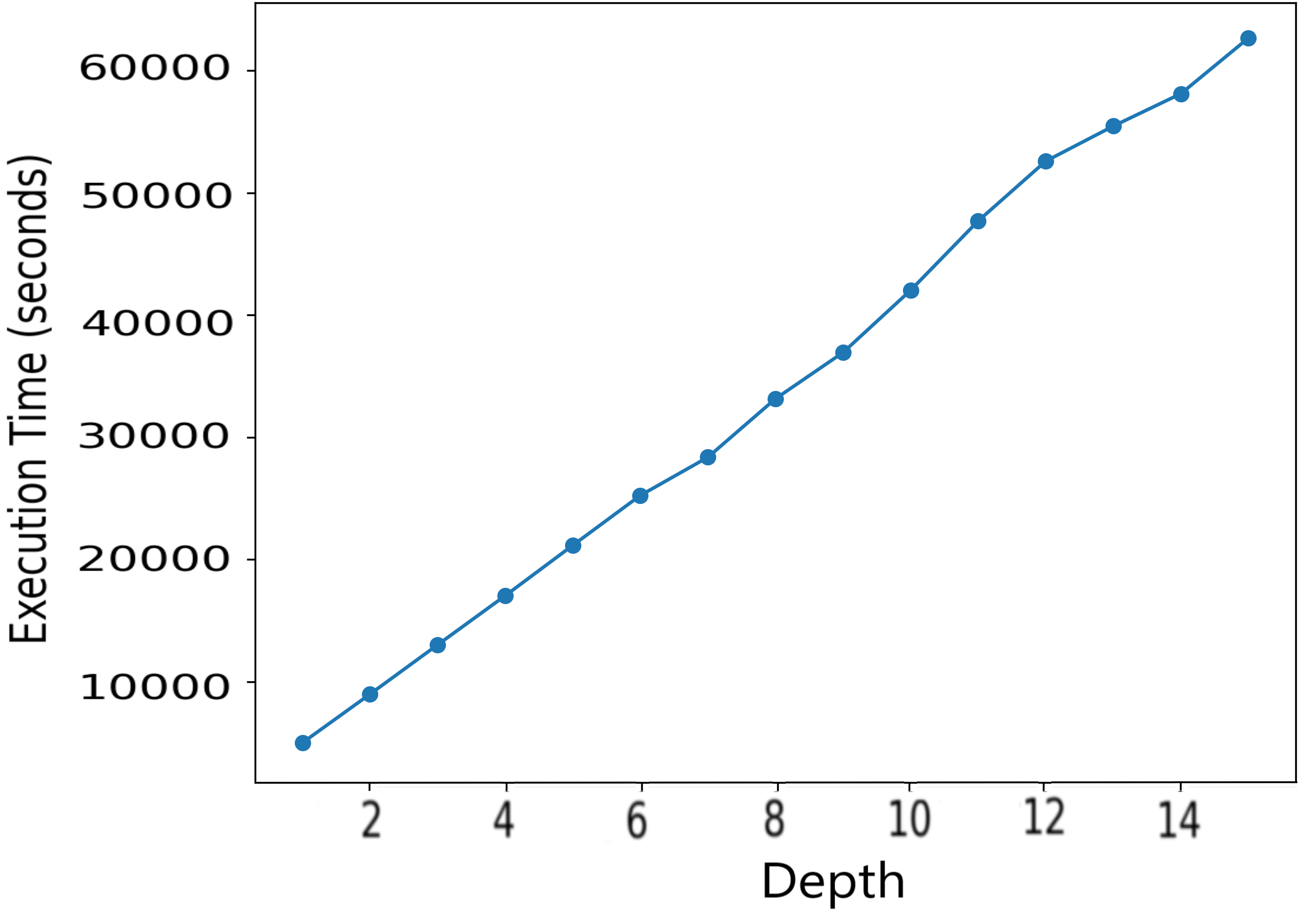}
}
\caption{Comparison of execution time, final loss and final accuracy as a function of circuit depth.}
\label{fig:final_metrics_depth}
\end{figure}

The following three graphs (Figure \ref{fig:final_metrics_depth}) respectively present the execution time, final lossand final accuracy as a function of the circuit depth.

 We observe a near linear increase in execution time with depth (Figure \ref{fig:execution_time_depth}), with the time exceeding 60,000 seconds for a depth of 14, which is approximately 16 hours. In terms of final loss (Figure \ref{fig:final_loss_depth}), the results show a marked improvement between depths 1 and 4, followed by stabilization from depth 5 onwards. Then, the final test accuracy (Figure \ref{fig:final_acc_depth}) follows a similar trend, plateauing around a depth of 4 with a final accuracy close to 0.98.\\
 
These results show that while increasing the circuit depth beyond 4 yields only marginal performance improvements, it comes at a significant cost in execution time. Therefore, it is essential to find a balance between circuit depth and computational efficiency for practical applications.\\

These results are consistent with findings from the literature, such as \cite{perez2019data}, which show that increasing the number of re-uploading layers can enhance model expressivity. Importantly, this highlights that expressivity is not simply proportional to the Hilbert space dimension or circuit depth. Instead, it depends on multiple factors including layer depth, gate structure, and the ability of parameter optimization to exploit the non-linear transformations introduced by data re-uploading:

\begin{equation}
\text{Expressivity} \sim g(L) \cdot h(U(\phi, \mathbf{x}), \mathbf{\theta}),
\end{equation}

where \( g(L) \) is a non-linear function that accounts for the diminishing returns of the increasing number of layers we re-upload \( L \), \( U(\phi, \mathbf{x}) \) is the parameterized unitary gates encoding the data and \( \mathbf{\theta} \) represents the tunable parameters of the circuit. Here \( \dim(\mathcal{H}) = 2 \) as it is for a single qubit and the expressivity is dominated primarily by the competition between the number of re-uploaded layers \( L \) and the gate structure. \\

As seen in our experiments, while the performance improves significantly for depths up to 4, further increases in depth provide only marginal gains. This can be explained by the bias-variance tradeoff \cite{briscoe2011conceptual,geman1992neural}: as we increase the depth, the model gains expressivity and reduces bias, but the improvements become smaller. This is likely because our classification problem is relatively simple and the additional complexity introduced by deeper circuits does not translate into significant performance improvements. The expressivity gained from re-uploading becomes less impactful once the model complexity exceeds the requirements of the task. Therefore, there is a practical upper bound to the benefits of increasing depth in this case.\\

Having examined the impact of circuit depth on the QRU's performance, we now turn our attention to the effect of the input normalization.

\paragraph{Input normalization}\hfill

\vspace{0.5cm}

In this section, we analyze the effect of input normalization on the performance of the QRU model. Specifically, we compare two normalization ranges: \( 0 \text{ to } 2\pi \) and \( -\pi \text{ to } \pi \). The results are presented in terms of final loss, final test accuracyand trainability.

\begin{figure}[H]
\centering
\includegraphics[width=0.8\textwidth]{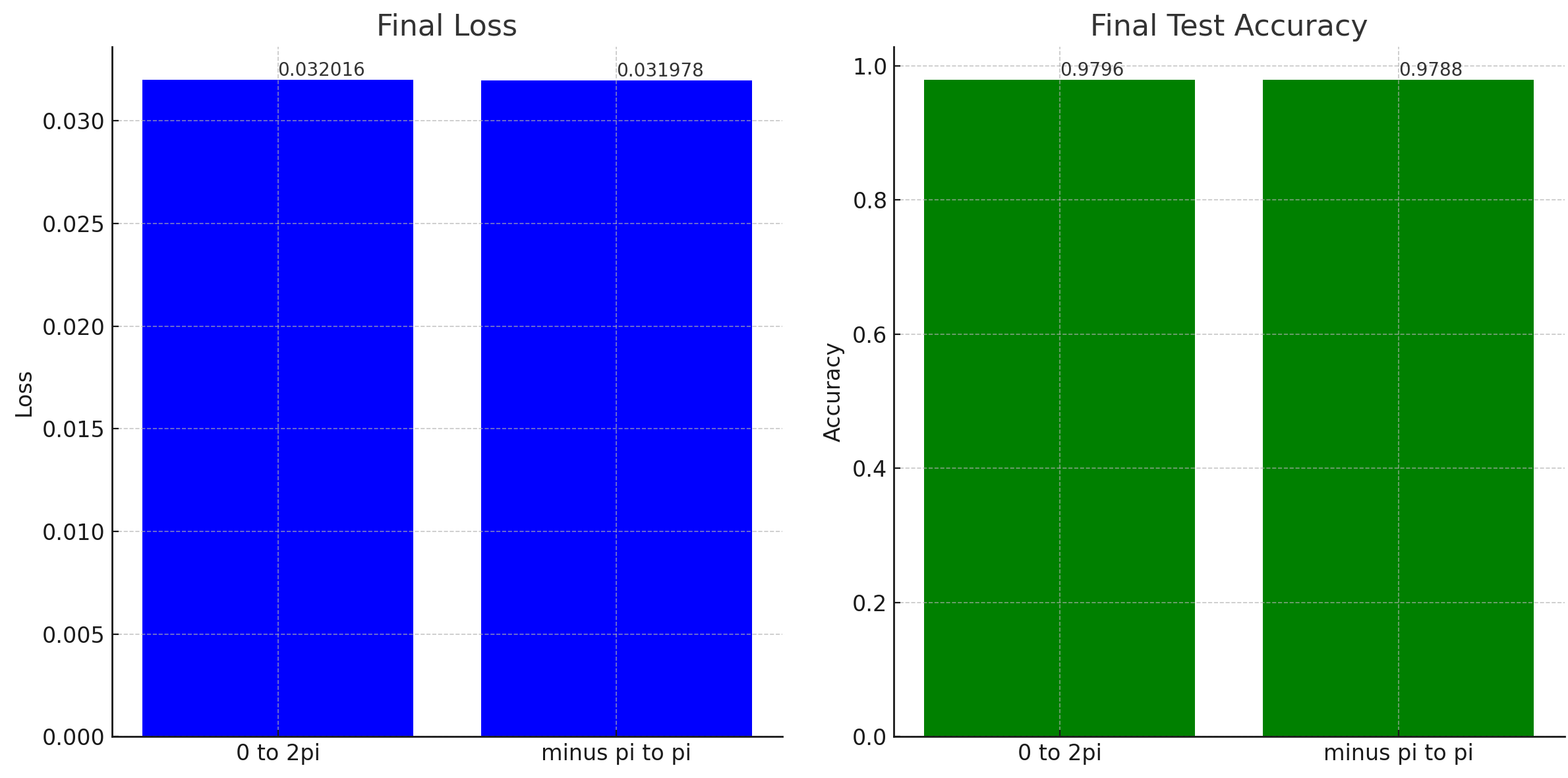}
\caption{Comparison of final loss and final test accuracy for input normalization ranges \( 0 \text{ to } 2\pi \) and \( -\pi \text{ to } \pi \)}
\label{fig:normalization_performance}
\end{figure}

As shown in Figure \ref{fig:normalization_performance}, there is no significant difference in performance between the two normalization ranges. The final mean loss is 0.032016 for \( 0 \text{ to } 2\pi \) and 0.031978 for \( -\pi \text{ to } \pi \), indicating virtually identical outcomes. Similarly, the final test accuracy is 0.9796 for \( 0 \text{ to } 2\pi \) and 0.9788 for \( -\pi \text{ to } \pi \), again showing no meaningful discrepancy. \\

\begin{figure}[H]
\centering
\includegraphics[width=\textwidth]{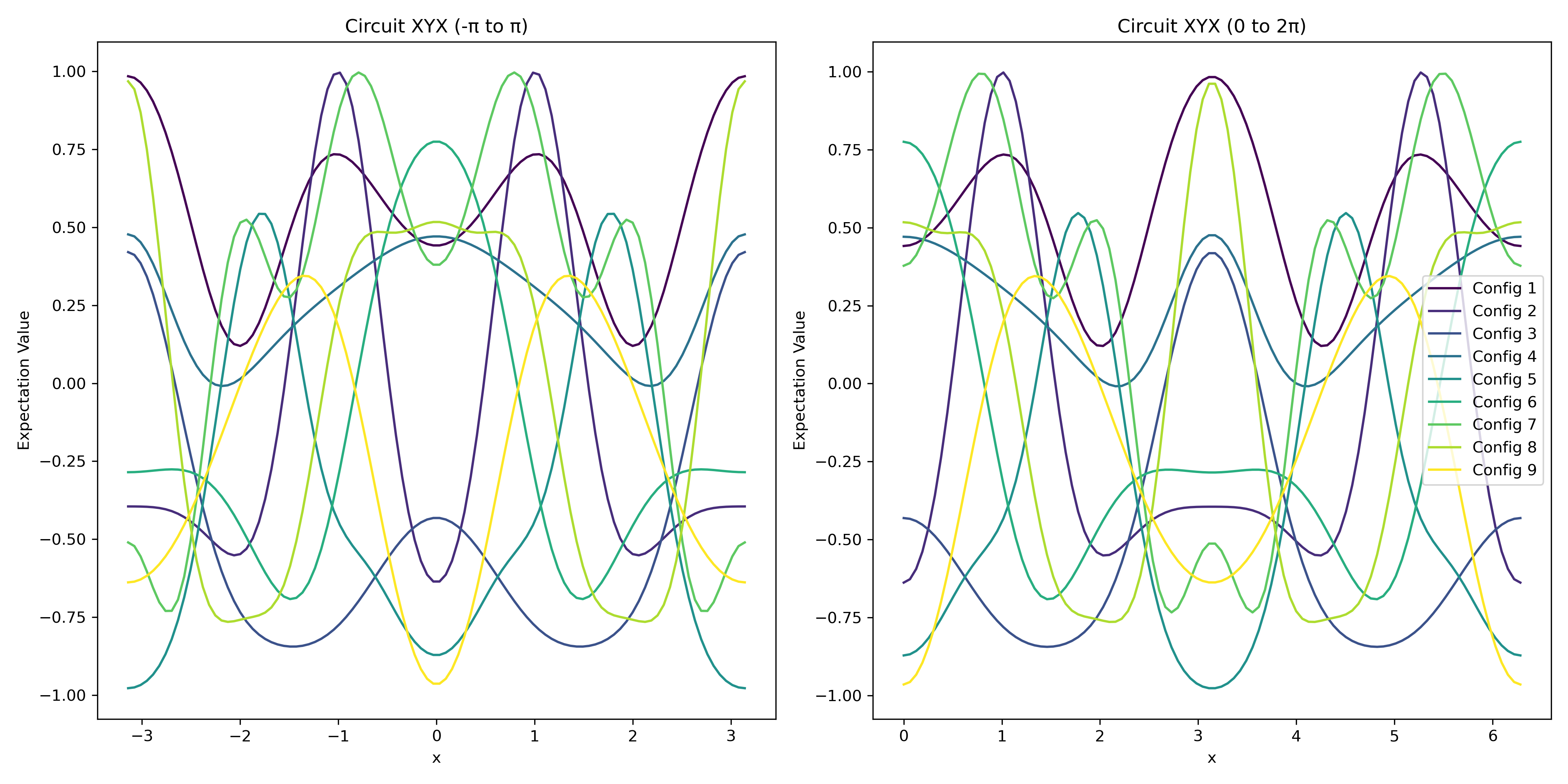}
\caption{Hypothesis functions $h_\theta(x)$ for 9 configurations of \( \theta \) for the circuit \( R_x - R_y - R_x \) over the intervals \( -\pi \text{ to } \pi \) (left) and \( 0 \text{ to } 2\pi \) (right) with a depth of 10}
\label{fig:fit_comparison}
\end{figure}

The lack of a significant difference can be attributed to the fact that our quantum circuits \( \gamma \), which govern the behavior of the QRU, are symmetric functions of the input over a period of \( 2\pi \). This symmetry is highlighted in the circuit fits shown in Figure \ref{fig:fit_comparison}, where we observe similar patterns in the behavior of the circuit for 9 different configurations of \( \theta \) over both normalization intervals.\\

Hypothesis functions \citep{barthe2023gradients} generated by QRU models are defined as the expectation value of a measurement observable \( Z \) applied to the output quantum state of the circuit. Mathematically, they are given by:

\begin{equation}
h_\theta(x) = \langle 0 | U^\dagger(\theta, x) Z U(\theta, x) | 0 \rangle,
\end{equation}

where \( U(\theta, x) \) is the parameterized quantum circuit that incorporates data encoding through iterative layers of single-qubit rotations. These hypothesis functions can be further expressed as generalized trigonometric polynomials:

\begin{equation}
h_\theta(x) = \sum_{\omega \in \Omega} a_\omega(\theta) e^{i \omega x},
\end{equation}

where \( \Omega \) is the set of available frequencies determined by the data encoding scheme and \( a_\omega(\theta) \) are Fourier coefficients dependent on the trainable parameters \( \theta \). 

This graph shows that shifting the normalization range by \( \pi \) to the right results in fit functions that remain even functions, exhibiting similar behavior to those normalized to the left of \( \pi \). As such, the expressivity does not change, which explains the lack of noticeable differences between the two ranges.\\

\begin{figure}[H]
\centering
\includegraphics[width=0.9\textwidth]{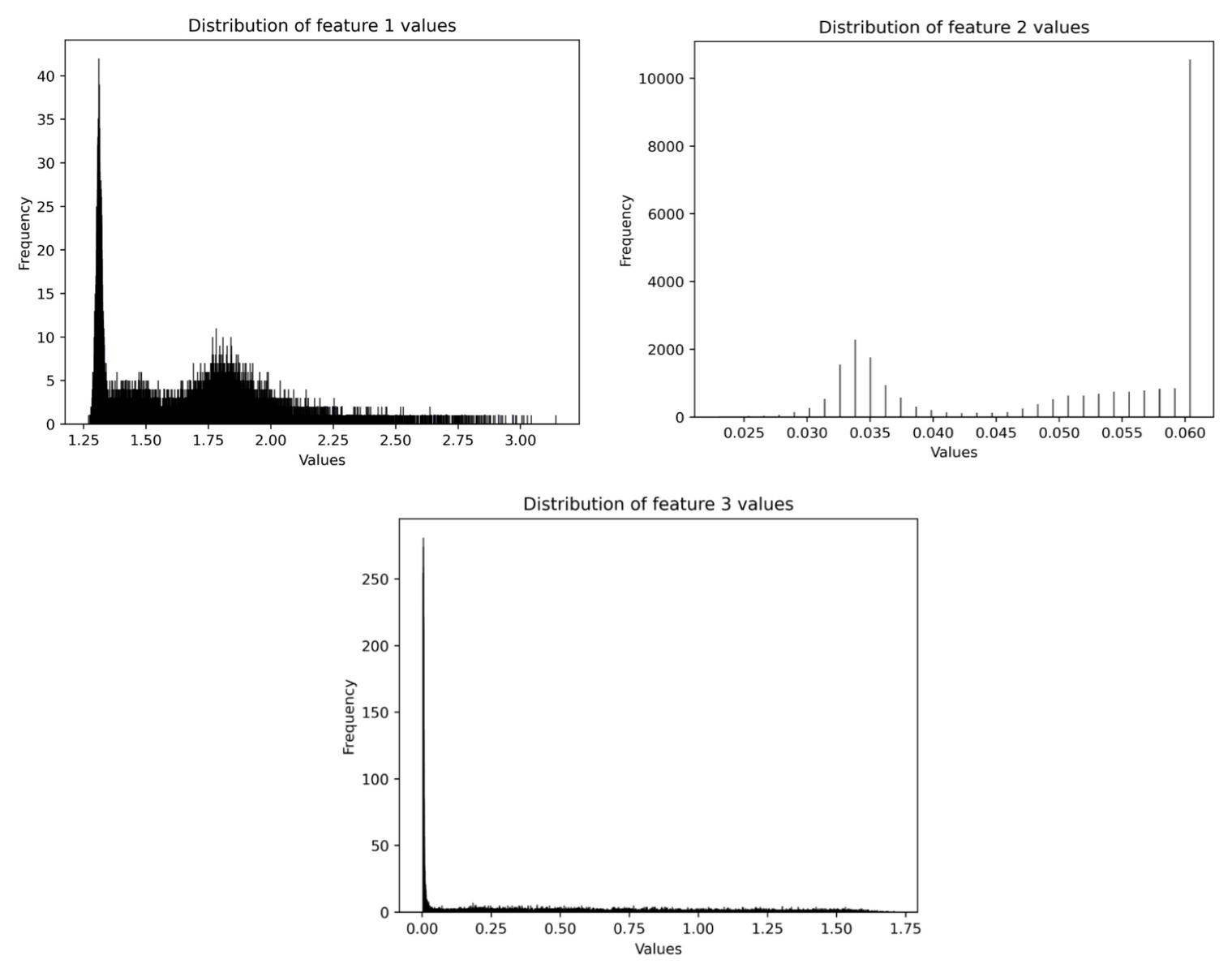}
\caption{Feature distributions}
\label{fig:non_normalized_vs_normalized}
\end{figure}

Previously, we tested the QRU model on non-normalized data and observed that it failed to learn properly. This behavior can be attributed to the distribution of input features (see Figure~\ref{fig:non_normalized_vs_normalized}): 
\begin{itemize}
    \item Feature~3 contains many zeros, but also a significant number of values in the range $0 < x \leq 1.75$; 
    \item Feature~1 shows a similar pattern, shifted upward by about~1.25, with many values around~1.125 and several in the interval $1.25 < x \leq 3.0$. 
    \item Feature~2 differs markedly, exhibiting no zeros and consistently small values, with a mode around~0.061 
\end{itemize}

This can be explained by considering the quantum state of the qubit: when the input values are very small, the qubit undergoes minimal rotations on the Bloch sphere, making it difficult to differentiate between states when the quantum measurement is performed along the Z-axis.

\paragraph{Rotation gates}\hfill

\vspace{0.5cm}

In this experiment, we analyzed the impact of different types of quantum rotations on the performance of the QRU model. We compared six circuits based on successive rotations of types $R_x$, $R_y$ and $R_z$, with the following structure:\\

\noindent \scriptsize $R_x - R_y - R_x :\hspace{0.5cm} \Qcircuit @C=0.2em @R=0.2em {\lstick{\ket{q}}& \gate{R_X(\theta_i)} & \gate{R_Y(\theta_{i+1} \times x_j)} & \gate{R_X(\theta_{i+2})} & \meter & \cw & \rstick{c} \\ } $\\

\noindent \scriptsize $R_x - R_z - R_x:\hspace{0.5cm} \Qcircuit @C=0.2em @R=0.2em{
      \lstick{\ket{q}} & \gate{R_X(\theta_i)} & \gate{R_Z(\theta_{i+1} \times x_j)} & \gate{R_X(\theta_{i+2})} & \meter & \cw & \rstick{c} \\
    }
    $\\
    
\noindent \scriptsize $R_y - R_x - R_y:\hspace{0.5cm} \Qcircuit @C=0.2em @R=0.2em{
      \lstick{\ket{q}} & \gate{R_Y(\theta_i)} & \gate{R_X(\theta_{i+1} \times x_j)} & \gate{R_Y(\theta_{i+2})} & \meter & \cw & \rstick{c} \\
    }
    $\\
    
\noindent \scriptsize $R_y - R_z - R_y: \hspace{0.5cm} \Qcircuit @C=0.2em @R=0.2em{
      \lstick{\ket{q}} & \gate{R_Y(\theta_i)} & \gate{R_Z(\theta_{i+1} \times x_j)} & \gate{R_Y(\theta_{i+2})} & \meter & \cw & \rstick{c} \\
    }
    $\\
    
\noindent \scriptsize $R_z - R_x - R_z:\hspace{0.5cm} \Qcircuit @C=0.2em @R=0.2em {
      \lstick{\ket{q}} & \gate{R_Z(\theta_i)} & \gate{R_X(\theta_{i+1} \times x_j)} & \gate{R_Z(\theta_{i+2})} & \meter & \cw & \rstick{c} \\
    }
    $\\
    
\noindent \scriptsize $R_z - R_y - R_z:\hspace{0.5cm} \Qcircuit @C=0.2em @R=0.2em{
      \lstick{\ket{q}} & \gate{R_Z(\theta_i)} & \gate{R_Y(\theta_{i+1} \times x_j)} & \gate{R_Z(\theta_{i+2})} & \meter & \cw & \rstick{c} \\
    }
    $\\

\normalsize
We choose to examine the performance of each of the three Euler rotations on the Bloch sphere for feature encoding, in both cases where the data-encoding rotation is surrounded by the other two Euler rotations as trainable parameters. Since we consider "sandwich" circuits, symmetric architectures rather than the asymmetric "triplet" type—we analyze the following six circuit configurations in our study. The performance of each type of circuit was evaluated in terms of final loss and final test accuracy. Figure \ref{fig:rotational_gates_performance} below illustrates the results obtained for these metrics.

\begin{figure}[H]
\centering
\includegraphics[width=0.9\textwidth]{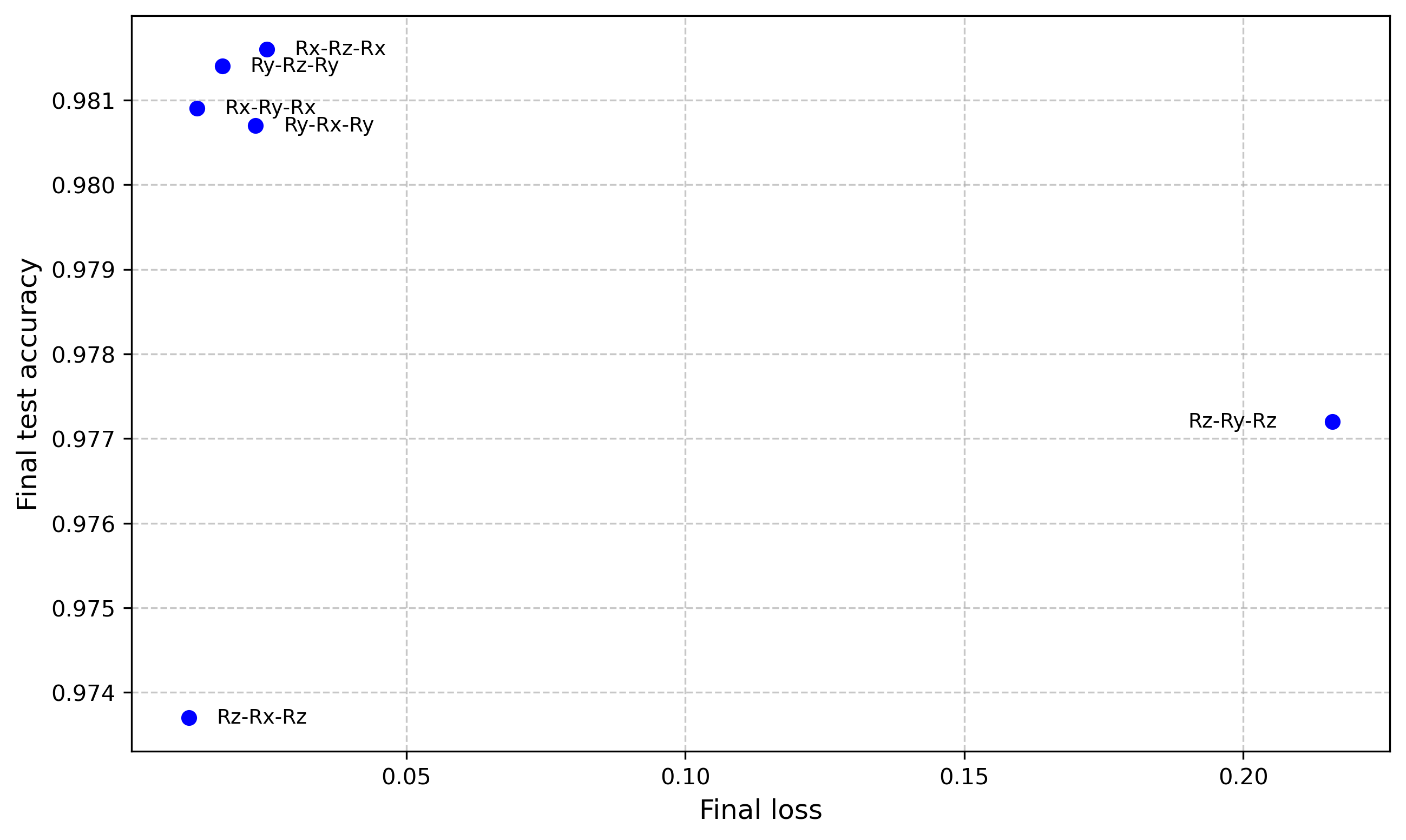}
\caption{Performance of different circuits based on rotation types}
\label{fig:rotational_gates_performance}
\end{figure}

The results show that the $R_x - R_y - R_x$ circuit achieves the best performance in terms of final loss, indicating that this combination of rotations allows for better optimization of the model. On the other hand, the $R_z - R_y - R_z$ circuit has the lowest final test accuracy, significantly lower than the other circuits. Conversely, the $R_x - R_z - R_x$ circuit and the $R_y - R_x - R_y$ circuit achieve high accuracy, with relatively low loss, showing that these rotation configurations offer a good balance between accuracy and stability. However, the $R_y - R_z - R_y$ circuit shows a higher final loss, suggesting that this type of rotation may introduce larger fluctuations in parameter optimization. Overall, the results indicate that fully parametrized rotations involving only $R_x$ and $R_y$ are more stable and perform better than those involving $R_z$ for the fully parametrized rotations.\\

Hubregtsen et al. (2021) \cite{hubregtsen2021evaluation} explored the correlation between the capacity of parameterized quantum circuits (PQC) to explore the Hilbert space and classification accuracy. They define a measure of expressivity, which quantifies the ability of a PQC to uniformly explore the Hilbert space, through the Kullback-Leibler divergence (DKL) between the fidelity distribution of the quantum states generated by the PQC and that of random Haar states. The expressivity is given by the following equation:

\begin{equation}
\text{Expr} = D_{\text{KL}}(P_{\text{PQC}}(F ; \Theta) || P_{\text{Haar}}(F)),
\end{equation}

where $P_{\text{PQC}}$ is the fidelity distribution of the PQC and $P_{\text{Haar}}$ is that of the random Haar states. The article concludes that rotations \(R_X\), \(R_Y\)and \(R_Z\) should be used together to maximize expressivity and therefore accuracy. None of these rotations is intrinsically superior, but their combination allows full exploration of the Hilbert space. This is directly related to the following theorem:

\begin{theorem}[Euler's theorem on rotations]\hfill \break

In a three-dimensional space, any rotation can be represented by a combination of elementary rotations around the $x$, $y$and $z$ axes, respectively denoted as $R_x$, $R_y$and $R_z$. More precisely, any rotation can be expressed as a single rotation by an angle $\theta$ around a fixed axis, according to the following decomposition:

\begin{equation}
    R(\theta, \hat{n}) = R_z(\phi) R_y(\Theta) R_x(\psi)
\end{equation}

where $\theta$ is the total rotation angle and $\hat{n}$ is the unit vector describing the rotation axis.

\end{theorem}

Nevertheless, our results show that the combination of $R_x$ and $R_y$ gates in quantum data re-uploading circuits tends to offer better performance in terms of accuracy and loss minimization compared to $R_x$. This may simply arise from the fact that during a measurement, we project our quantum state onto the Z-axis and therefore, rotating around the Z-axis does not change the projection of the qubit on the Z-axis.\\

\begin{figure}[H]
\centering
\includegraphics[width=1.0\textwidth]{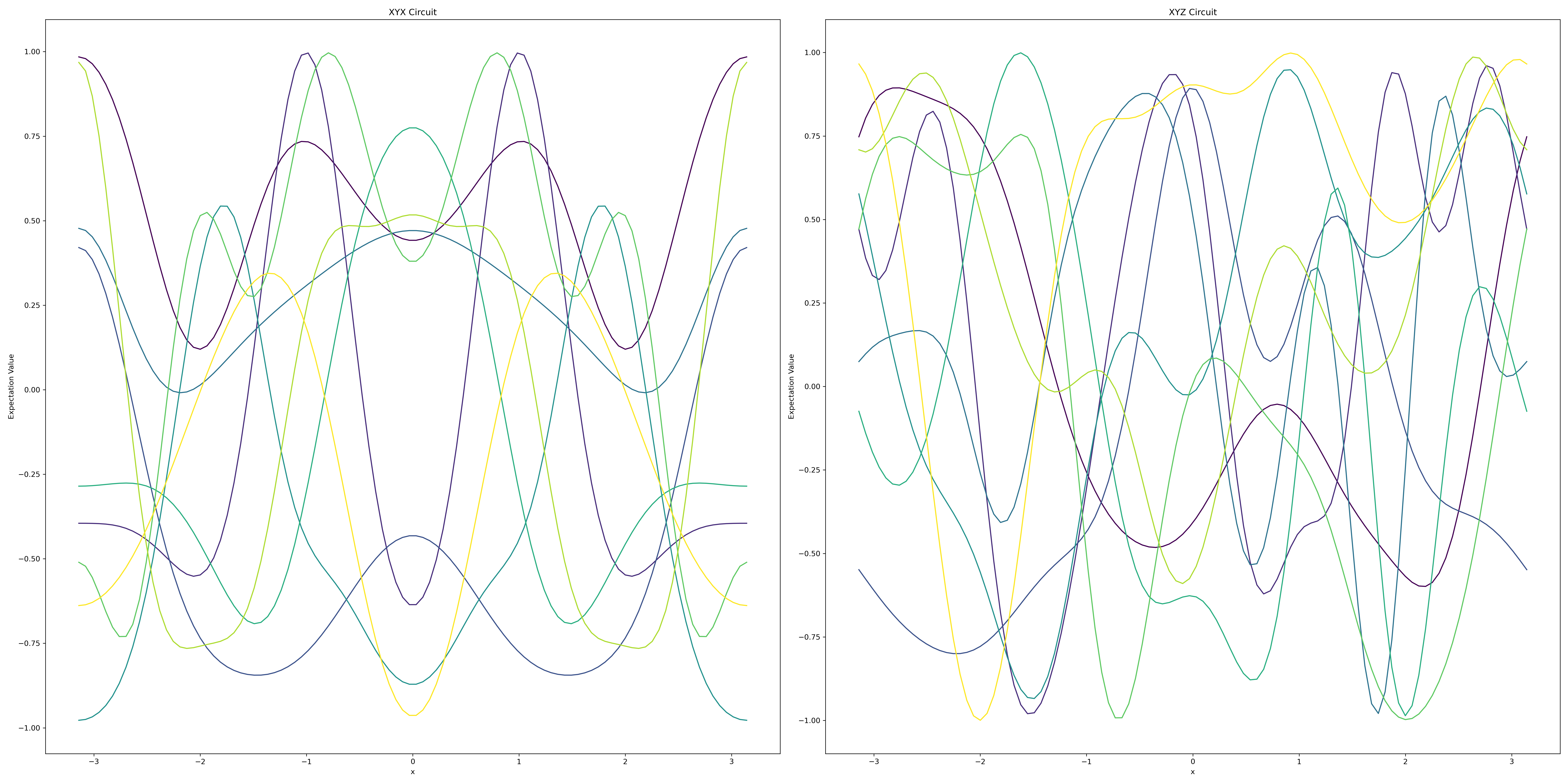}
\caption{Hypothesis functions $h_\theta(x)$ for 9 configurations of \( \theta \) for the circuit \( R_x - R_y - R_x \) (left) and  \( R_x - R_y - R_z \) (right). }
\label{fig:sand_tripl}
\end{figure}

This pushes us to test other types of rotations, which we could call "triplets" ($R_x - R_y - R_z$), involving rotations around all three axes, rather than using simpler "sandwich" rotations like in our case.  We plotted the QRU functions for the same depth and the same combination of theta parameters for a "sandwich" circuit and a "triplet" circuit. These results are presented in Figure \ref{fig:sand_tripl}. 

\begin{figure}[H]
\centering
\includegraphics[width=1.0\textwidth]{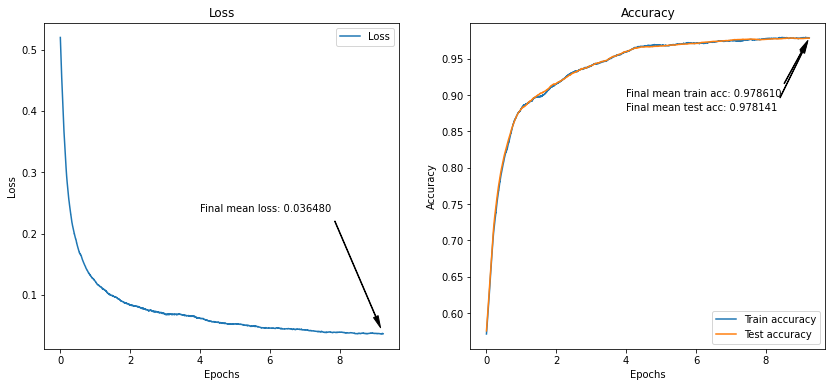}
\caption{Performance of the "triplet" quantum circuit}
\label{fig:triplet}
\end{figure}

We observe that "sandwich" circuits are constrained to be even functions of the input, whereas "triplet" circuits have more flexibility. Theoretically, this should result in better expressivity.
Figure \ref{fig:triplet} shows the results obtained with  \( R_x - R_y - R_z \) quantum circuit.

The final mean loss is 0.036480, with a final mean train accuracy of 0.978610 and test accuracy of 0.978141. Comparing this with the "sandwich" circuits using only two rotations, such as $R_x - R_y - R_x$ or $R_x - R_z - R_x$, where the final accuracies were slightly higher and with lower final losses, it becomes evident that the two-rotation circuits tend to outperform in minimizing loss in our case. While three-rotation circuits like $R_x - R_y - R_z$ theoretically offer higher expressivity, in this specific case, the increased complexity does not significantly improve accuracy, but instead results in a slightly higher loss. This suggests that two-rotation circuits provide a simpler yet equally effective approach for our type of classification tasks, balancing between expressivity and optimization stability.\\

Next, we analyze how the number of trainable parameters per input affects the model's performance.

\paragraph{Number of trainable parameters per input}\hfill

\vspace{0.5cm}

In this part, we analyze the impact of varying the number of parameters \( \theta \) per input on the performance of the QRU model. We tested five different quantum circuit configurations, each with an increasing number of parameters per input, as described below:

\vspace{0.2cm}

 1 parameter \( \theta \) per input:

\vspace{0.1cm}

\scriptsize \( ~~~~ \Qcircuit @C=0.2em @R=0.2em {
      \lstick{\ket{q}} & \gate{R_X(\theta_i)} & \gate{R_Y(x_j)}   }\)

\vspace{0.2cm}

\normalsize 2 parameters \( \theta \) per input:

\vspace{0.1cm}

\scriptsize \( ~~~~ \Qcircuit @C=0.2em @R=0.2em {
      \lstick{\ket{q}} & \gate{R_X(\theta_i)} & \gate{R_Y(x_j)} & \gate{R_X(\theta_{i+1})}  }\)

\vspace{0.2cm}

\normalsize 3 parameters \( \theta \) per input:

\vspace{0.1cm}

\scriptsize  \(~~~~ \Qcircuit @C=0.2em @R=0.2em {
      \lstick{\ket{q}} & \gate{R_X(\theta_i)} & \gate{R_Y(\theta_{i+1} \times x_j)} & \gate{R_X(\theta_{i+2})}  }\)

\vspace{0.2cm}

\normalsize 4 parameters \( \theta \) per input:

\vspace{0.1cm}

\scriptsize  \( ~~~~ \Qcircuit @C=0.2em @R=0.2em {
      \lstick{\ket{q}} & \gate{R_X(\theta_i)} & \gate{R_Y(\theta_{i+1} \times x_j + \theta_{i+2})} & \gate{R_X(\theta_{i+3})}  }\)

\vspace{0.2cm}

\normalsize 5 parameters \( \theta \) per input:

\vspace{0.1cm}

\scriptsize \( ~~~~ \Qcircuit @C=0.2em @R=0.2em {\lstick{\ket{q}} & \gate{R_X(\theta_i)} & \gate{R_Y(\theta_{i+1}^2 \times x_j + \theta_{i+2} \times x_j + \theta_{i+3})} & \gate{R_X(\theta_{i+4})}  }\)\\

\vspace{0.2cm}

\begin{figure}[H]
\centering
\includegraphics[width=1.0\textwidth]{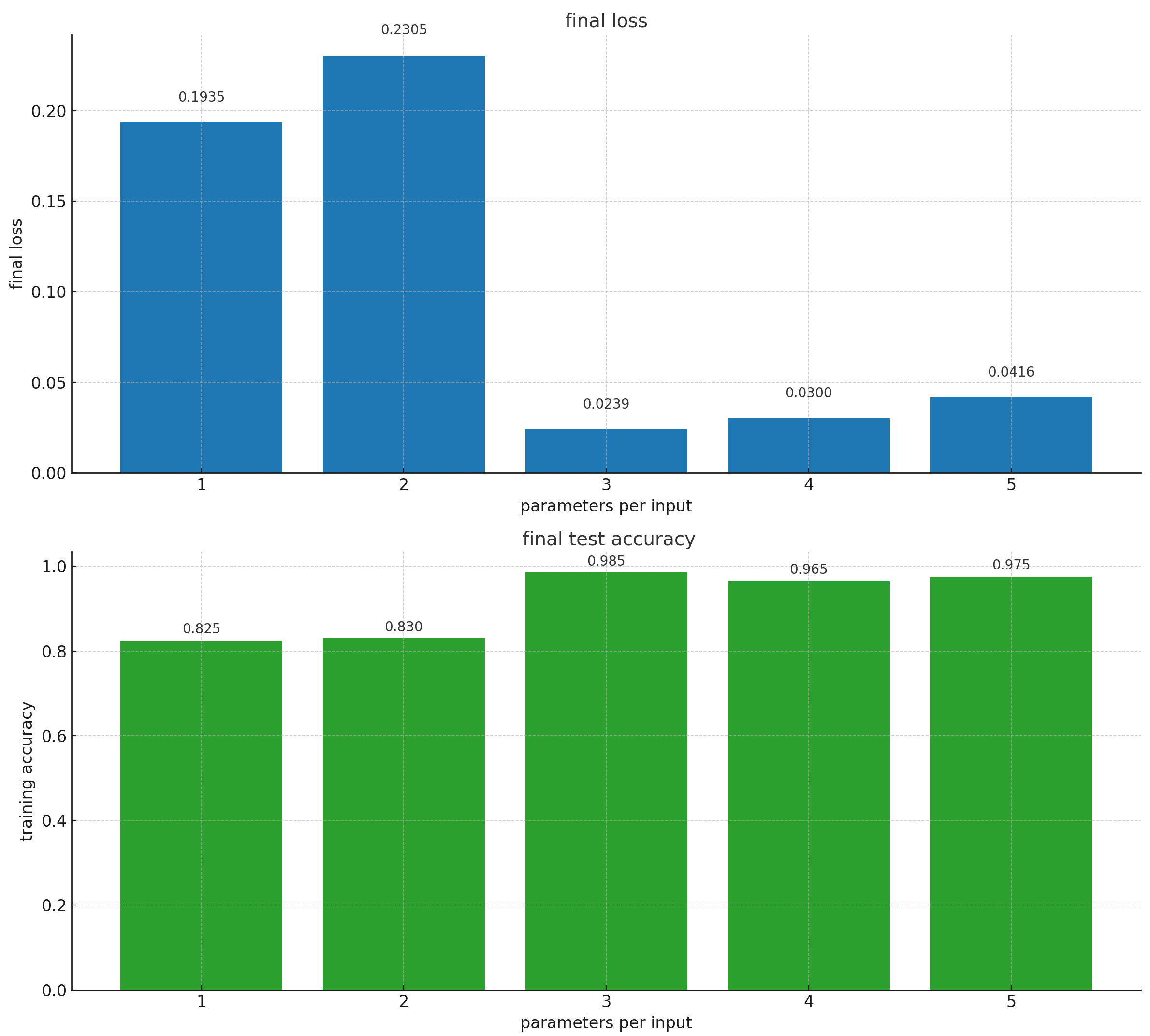}
\caption{Performance of the QRU model based on the number of parameters per input. Top: final loss; Bottom: final test accuracy.}
\label{fig:params_per_input_performance}
\end{figure}

\normalsize The results of these tests are shown in Figure \ref{fig:params_per_input_performance}.

As seen in Figure \ref{fig:params_per_input_performance}, the final loss decreases dramatically when moving from 1 or 2 parameters per input (0.1935 and 0.2305, respectively) to 3 parameters, where the final loss drops to 0.0239. However, after 3 parameters per input, the final loss slightly increases again for 4 and 5 parameters (0.0300 and 0.0416, respectively). This indicates that increasing the number of parameters per input beyond 3 does not significantly improve loss minimization and may even complicate optimization.\\

Similarly, the final test accuracy improves from 0.825 with 1 parameter to 0.985 with 3 parameters. However, as with the loss, the accuracy plateaus and slightly decreases with 4 and 5 parameters (0.965 and 0.975). This suggests that 3 parameters per input represent the optimal configuration for balancing model complexity and performance in terms of both accuracy and loss. In summary, while optimizing the number of parameters per input improves the expressivity and performance of the quantum circuit, it also increases the computational complexity. In the next section, we explore the use of GPU acceleration to reduce execution time of the QRU model.\\

After evaluating the QRU model's performance based on individual hyperparameter variations, we now move to a more comprehensive approach by considering the optimization of all hyperparameters simultaneously.

\subsubsection{Training hyperparameters}

Here, we explore the impact of training-related hyperparameters, including batch size, optimizer, loss function and learning rate, on model convergence, stability and performance.

\paragraph{Batch Size}\hfill

\vspace{0.5cm}

In our implementation, we do not vary the mini-batch size in the standard sense (i.e., evaluating a batched quantum circuit). Instead, we vary the optimizer update frequency through gradient accumulation. Each forward/backward pass processes a single example, and the optimizer is applied only once every $\texttt{update\_freq}$ samples. Concretely, gradients from $B_{\mathrm{eff}}=\texttt{update\_freq}$ individual samples are accumulated and averaged before the parameter update:
\[
\Delta\theta \;=\;
-\,\eta\,
\frac{1}{B_{\mathrm{eff}}}
\sum_{t=1}^{B_{\mathrm{eff}}}
\nabla_\theta \ell(x_t;\theta),
\qquad
\eta = 5\times10^{-5}
\]
With this averaging, the effective step magnitude is not artificially scaled with $B_{\mathrm{eff}}$. Therefore, changing $B_{\mathrm{eff}}$ primarily probes the effect of performing fewer (or more) parameter updates per epoch at fixed epoch budget, rather than an unintended learning-rate rescaling.\\

\begin{figure}[H]
\centering
\begin{subfigure}{0.48\textwidth}
    \centering
    \includegraphics[width=\textwidth]{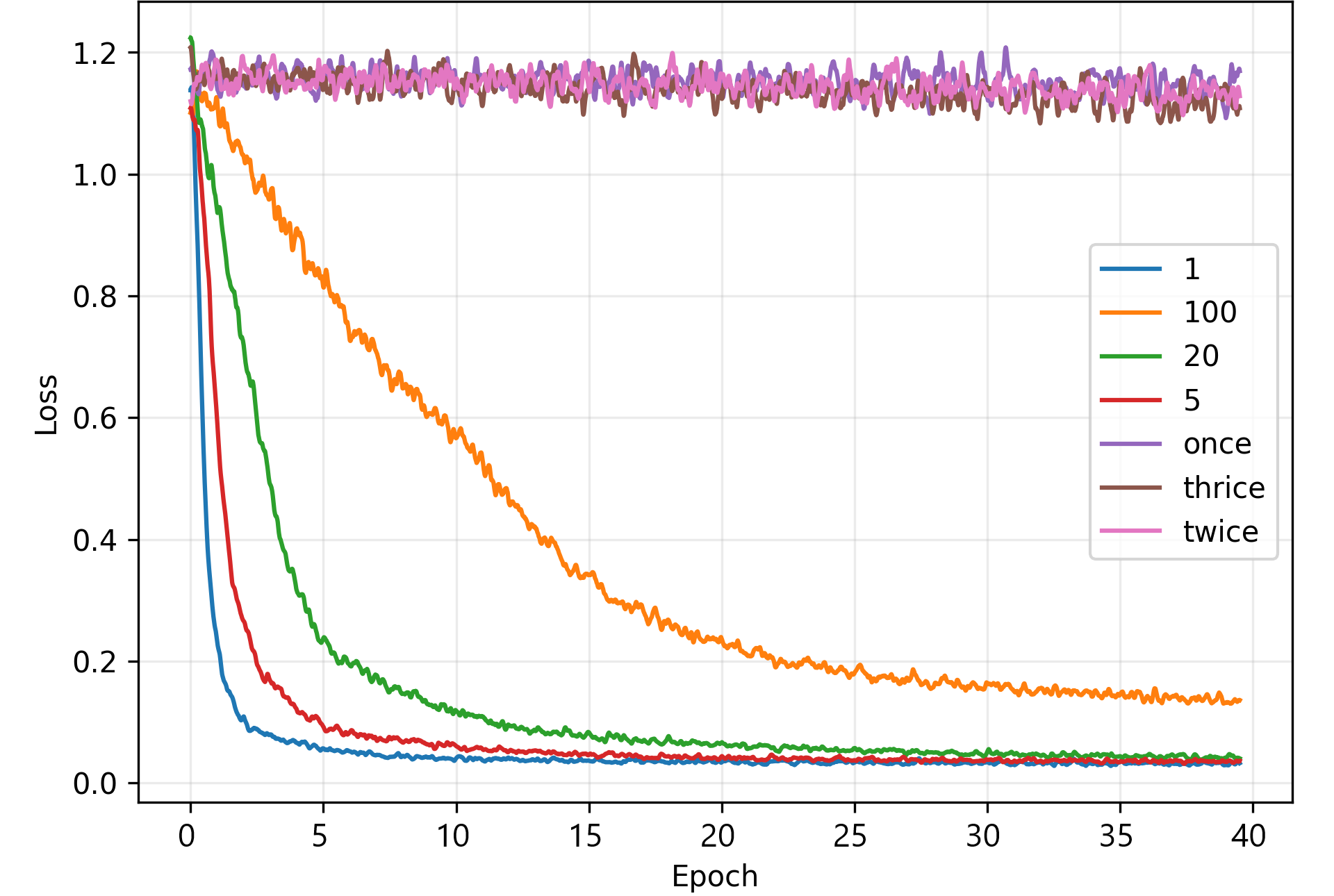}
    \caption{Training loss vs epoch.}
    \label{fig:loss_batch_size}
\end{subfigure}
\hfill
\begin{subfigure}{0.48\textwidth}
    \centering
    \includegraphics[width=\textwidth]{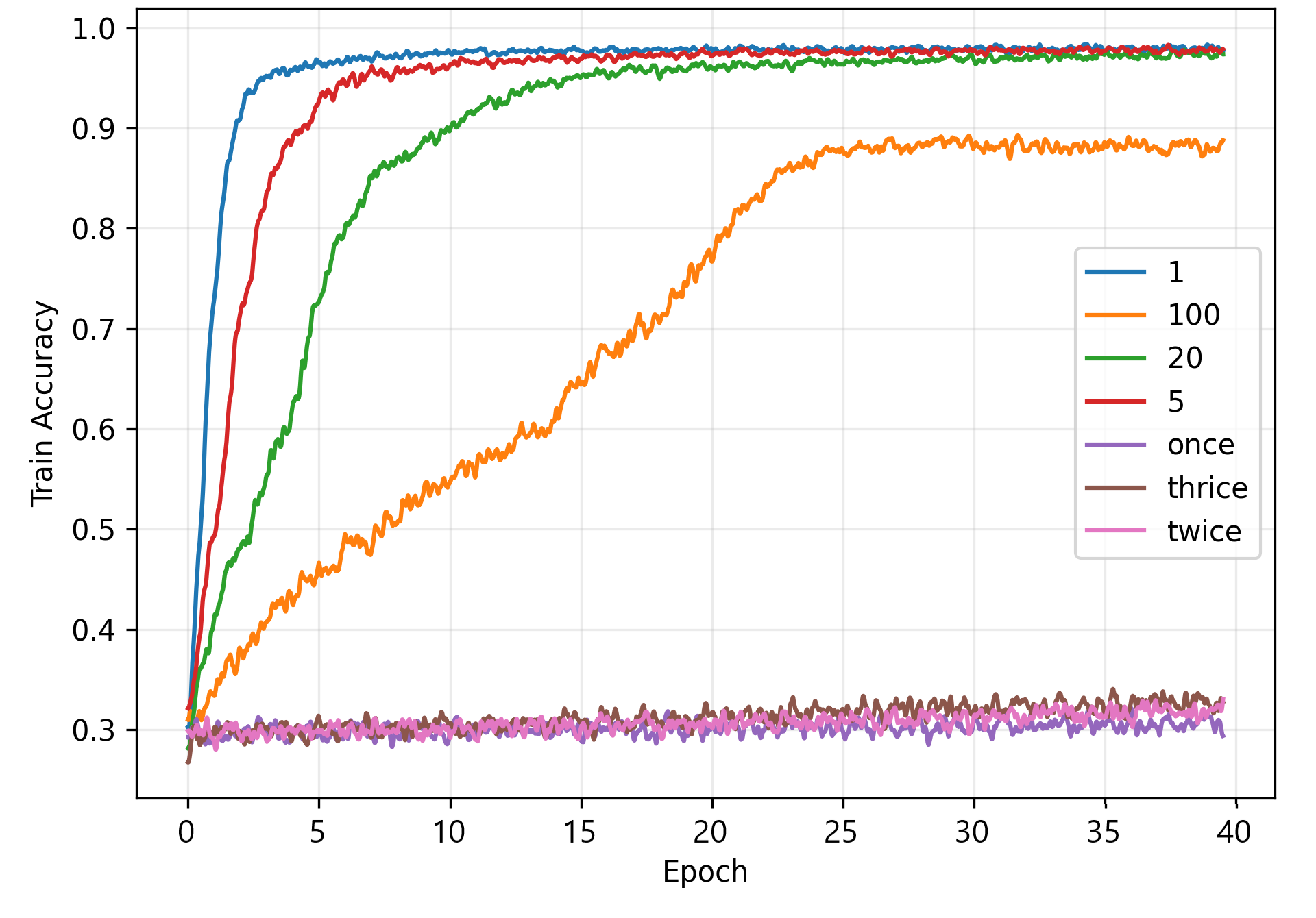}
    \caption{Training accuracy vs epoch.}
    \label{fig:accuracy_batch_size}
\end{subfigure}
\caption{Effect of the optimizer update interval $B_{\mathrm{eff}}$ (gradient accumulation with averaging) on training dynamics.}
\label{fig:accuracy_loss_batch_size}
\end{figure}

The curves in Fig.~\ref{fig:accuracy_loss_batch_size} and the metrics in Table~\ref{tab:batch_sizes} report results for $B_{\mathrm{eff}}\in\{1,5,20,100\}$, along with three extreme regimes corresponding to approximately one, two, or three optimizer updates per epoch (``once/twice/thrice''). Under a fixed number of epochs, increasing $B_{\mathrm{eff}}$ reduces the number of optimizer steps per epoch by a factor $\approx 1/B_{\mathrm{eff}}$, which directly impacts convergence. We observe that moderate update intervals remain effective (e.g., $B_{\mathrm{eff}}=20$ yields a final accuracy of $0.973$ and loss $0.042$), while very large update intervals significantly slow optimization and lead to worse final performance (e.g., $B_{\mathrm{eff}}=100$ yields $0.881$ accuracy and loss $0.136$). In the extreme ``once/twice/thrice'' settings (roughly $1$--$3$ optimizer steps per epoch), training fails to make progress and converges near chance-level accuracy ($\approx 0.31$--$0.33$), consistent with an insufficient number of parameter updates within the chosen epoch budget.

We report final train/evaluation accuracies and losses, as well as a trainability proxy defined as the area under the training-loss curve over epochs (AUC-loss):
\begin{equation}
\label{eq:trainability}
\text{Trainability (AUC-loss)} = \int_0^E \text{Loss}(e)\,de,
\end{equation}
where $E$ denotes the total number of training epochs. As used here, AUC-loss is computed from the recorded training-loss trace on the epoch axis; lower values indicate faster and steadier loss decay within the fixed epoch budget.

\begin{table}[H]
\centering
\small
\caption{Metrics for different update intervals $B_{\mathrm{eff}}=\texttt{update\_freq}$ (gradient accumulation with averaging).}
\label{tab:batch_sizes}
\setlength{\tabcolsep}{3pt}
\begin{tabular}{lccccc}
\toprule
$B_{\mathrm{eff}}$ & Train acc & Eval. acc & AUC-loss & Time (s) & Eval. loss \\
\midrule
1   & \textbf{0.979} & \textbf{0.980} & \textbf{1.328} & 231{,}569 & \textbf{0.032} \\
5   & 0.977 & 0.978 & 2.057 & 228{,}794 & 0.035 \\
20  & 0.973 & 0.973 & 3.858 & \textbf{222{,}340} & 0.042 \\
100 & 0.880 & 0.881 & 10.976 & 223{,}095 & 0.136 \\
\midrule
thrice ($\approx 3$ steps/epoch) & 0.328 & 0.328 & 29.474 & 230{,}494 & 1.125 \\
twice ($\approx 2$ steps/epoch)  & 0.318 & 0.318 & 30.180 & 228{,}776 & 1.130 \\
once  ($\approx 1$ step/epoch)   & 0.307 & 0.307 & 30.482 & 234{,}537 & 1.141 \\
\bottomrule
\end{tabular}
\end{table}

Overall, these results indicate that performance degradation at large $B_{\mathrm{eff}}$ is consistent with an optimization-budget effect: when epochs are held fixed, larger update intervals yield fewer optimizer steps and slower convergence. Importantly, because we average accumulated gradients, these trends should not be attributed to step-size inflation (overshooting). Moreover, runtime varies only modestly with $B_{\mathrm{eff}}$, which is expected in this setting where the dominant cost is the per-sample quantum circuit evaluation rather than the optimizer step itself. We now turn to the analysis of the optimizer, a key hyperparameter that influences convergence speed and final performance.

\paragraph{Optimizer}\hfill

\vspace{0.5cm}

We compare here the performance of different optimizers, including SGD, RMSProp and Adam, based on their impact on the performances of the QRU. Table~\ref{tab:optimizer_performance} summarizes the results of the three different optimizers.

\begin{table}[h]
\centering
\caption{Performance comparison of optimizers.}
\label{tab:optimizer_performance}
\setlength{\tabcolsep}{8pt}
\begin{tabular}{lcccc}
\toprule
Optimizer & Execution time (s) & Final test accuracy & Final loss & Trainability \\
\midrule
RMSprop & 142{,}569.70 & 0.844 & 0.109 & \textbf{0.181} \\
SGD     & \textbf{21{,}217.02} & 0.334 & 1.024 & 2.987 \\
Adam    & 143{,}880.52 & \textbf{0.984} & \textbf{0.0207} & 0.776 \\
\bottomrule
\end{tabular}
\end{table}

SGD (Stochastic Gradient Descent) is the simplest optimizer, updating parameters using only the gradient of the loss function. It calculates the gradient for each batch and updates the parameters as follows:

\begin{equation}
\theta_{t+1} = \theta_t - \eta \nabla_\theta J(\theta_t)
\end{equation}

where \(\theta_t\) is the model parameter at step \(t\), \(\eta\) is the learning rate and \(J(\theta_t)\) is the loss function.\\

RMSProp (Root Mean Square Propagation) adapts the learning rate by dividing it by a moving average of the square of the gradients. This helps RMSProp handle problems where the magnitude of the gradients varies greatly:

\begin{equation}
\theta_{t+1} = \theta_t - \frac{\eta}{\sqrt{v_t + \epsilon}} \nabla_\theta J(\theta_t)
\end{equation}

where \(v_t\) is the exponentially weighted moving average of the squared gradients and \(\epsilon\) is a small value to prevent division by zero.\\

Adam (Adaptive Moment Estimation) combines ideas from both RMSProp and momentum. It calculates the exponentially weighted average of both the gradients and the squared gradients:

\begin{equation}
m_t = \beta_1 m_{t-1} + (1 - \beta_1) g_t 
\end{equation}
\begin{equation*}
v_t = \beta_2 v_{t-1} + (1 - \beta_2) g_t^2
\end{equation*}
where \(m_t\) and \(v_t\) are the first and second moment estimates (similar to momentum), \(\beta_1\) and \(\beta_2\) are decay rates and \(g_t\) is the gradient. The parameter update is then given by:

\begin{equation}
\theta_{t+1} = \theta_t - \eta \frac{m_t}{\sqrt{v_t} + \epsilon}
\end{equation}

SGD shows the lowest runtime per epoch, as expected from its simpler update rule. However, with the learning rate used in our experiments it reached lower final accuracy and higher loss than RMSProp and Adam. This likely reflects the greater sensitivity of SGD to learning-rate choice, whereas adaptive methods such as RMSProp and Adam are more robust. RMSProp converges faster than SGD but underperformed Adam in final accuracy and loss. Adam achieved the best overall performance, at the cost of a slightly higher runtime due to the additional moment estimates.\\

\begin{figure}[H]
\centering
\includegraphics[width=1.0\textwidth]{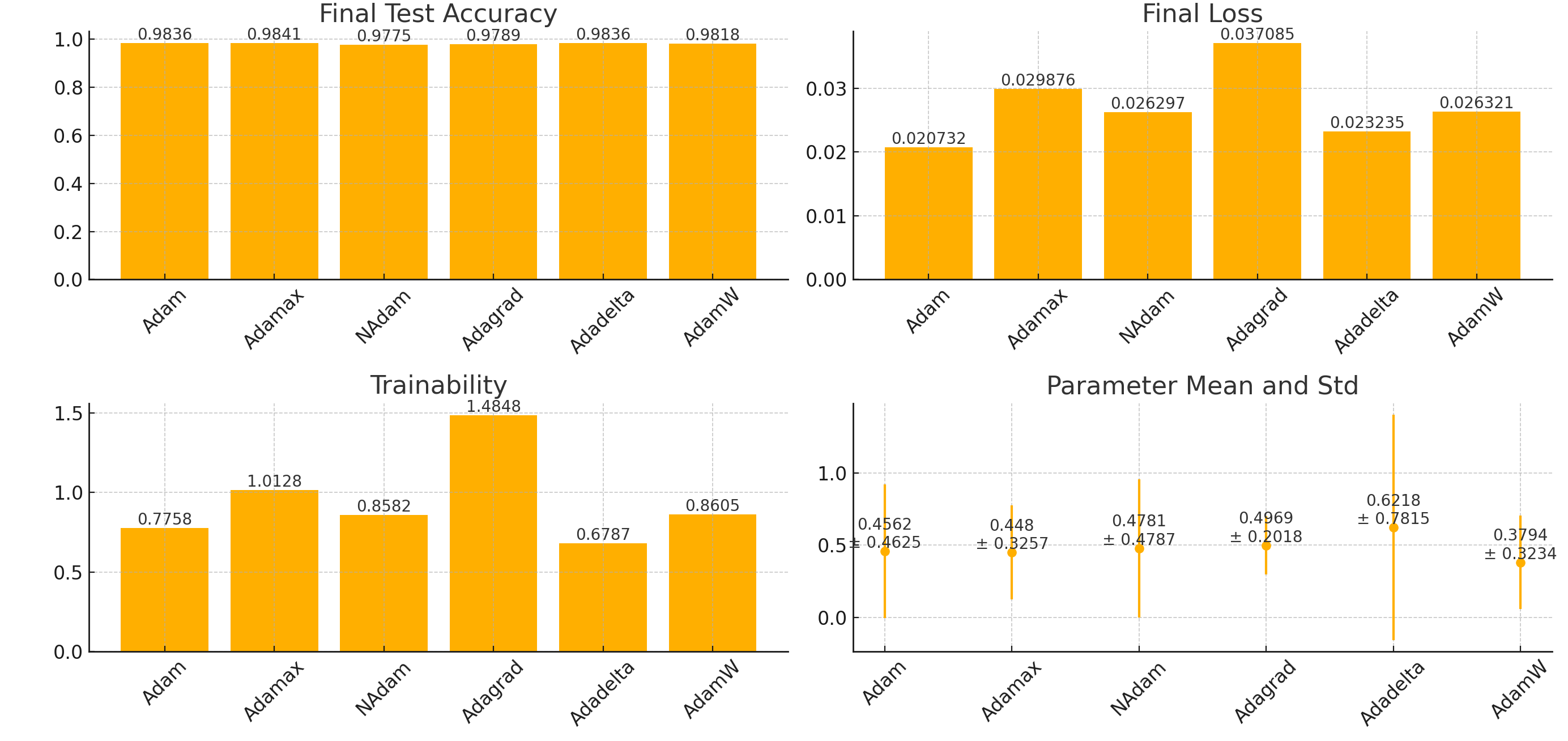}
\caption{Comparison of Adam-derived optimizers across metrics}
\label{fig:adam_variants}
\end{figure}

The figure below (Figure \ref{fig:adam_variants}) compares several optimizers derived from Adam, including Adamax, Nadam, Adagrad, Adadelta and AdamW.

Each Adam variant brings slight modifications to the original algorithm. Adamax uses the infinity norm (instead of the second norm), making it more stable in certain cases. Nadam introduces Nesterov momentum into Adam, allowing for faster convergence in some settings. Adagrad adapts the learning rate based on the history of gradients, making it suitable for sparse data but leading to slower learning in the long run. Adadelta dynamically adjusts the learning rate without needing an initial learning rate, making it more resilient to noisy data. And finally, AdamW modifies the weight decay method to achieve better generalization. \cite{loshchilov2019decoupledweightdecayregularization} As seen in the results, the performance of these optimizers is closely tied to the specific characteristics of the quantum circuit. Adagrad, for example, shows higher trainability so it suffers from slower learning, which is reflected in its higher final loss.\\

With the optimizer performance thoroughly analyzed, we now turn our attention to the loss function used for the training.

\paragraph{Loss Function}\hfill

\vspace{0.5cm}

We explored three different loss functions: L1, L2 and Huber. These are used to measure the discrepancy between predicted values $\hat{y}$ and actual values $y$. The L1 loss is defined as:

\begin{equation}
    L1(y, \hat{y}) = \sum_{i=1}^n |y_i - \hat{y}_i|
\end{equation}

It computes the sum of the absolute differences between predictions and true values, making it robust to outliers but slower to converge. The L2 loss, on the other hand, is given by:

\begin{equation}
    L2(y, \hat{y}) = \sum_{i=1}^n (y_i - \hat{y}_i)^2
\end{equation}

This function amplifies larger errors by squaring the differences, making it more sensitive to outliers. Lastly, the Huber loss \cite{huber1992robust} is a hybrid of L1 and L2:

\begin{equation}
    L_\delta(y, \hat{y}) = 
    \begin{cases} 
    \frac{1}{2} (y - \hat{y})^2 & \text{if } |y - \hat{y}| \leq \delta \\
    \delta (|y - \hat{y}| - \frac{1}{2} \delta) & \text{otherwise}
    \end{cases}
\end{equation}

Huber behaves like L2 for small errors and like L1 for large errors, offering a balance between handling outliers and efficient optimization. 

\begin{figure}[H]
\centering
\includegraphics[width=0.5\textwidth]{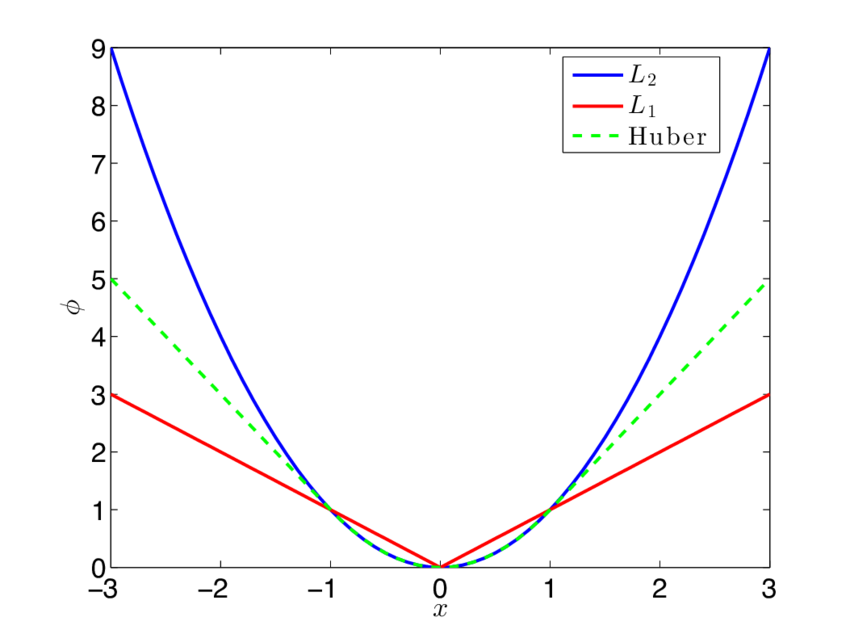}
\caption{Difference between the three loss functions used}
\label{fig:loss_functions}
\end{figure}

\begin{figure}[H]
\centering
\includegraphics[width=1.0\textwidth]{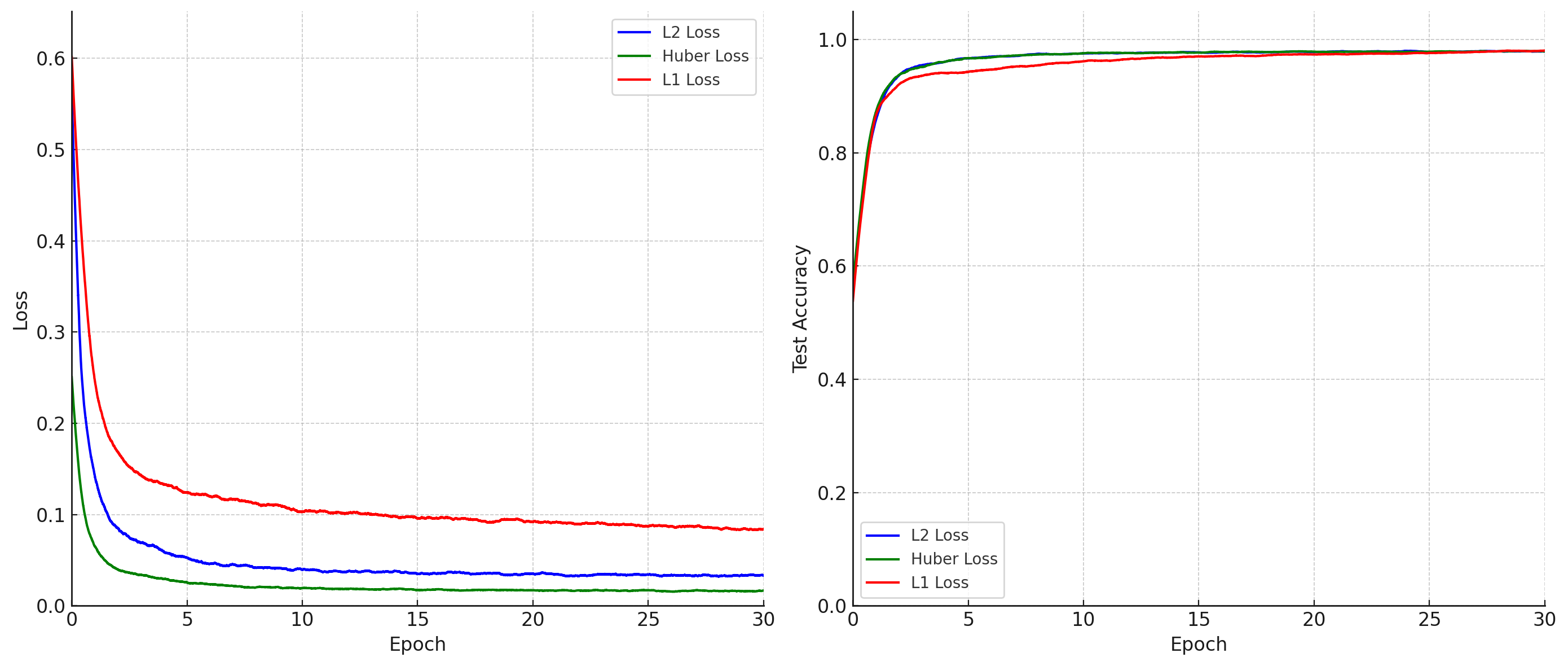}
\caption{Evolution of the loss and the test accuracy over the epochs with the three different loss function}
\label{fig:loss_test_acc_VS_epoch_loss_functions}
\end{figure}

The results of our experiments using these loss functions over 30 epochs are shown in the figure \ref{fig:loss_test_acc_VS_epoch_loss_functions}. We observe that while the final loss values differ significantly between the functions, the final test accuracy remains nearly the same for all three.

The difference in final loss values between the loss functions can be explained by the optimization properties of each function. For instance, the Huber loss, which combines aspects of both L1 and L2 losses, is particularly useful for handling outliers, making it less sensitive to quantum noise. This robustness allows for better minimization of the final loss.

\begin{figure}[H]
\centering
\includegraphics[width=1.0\textwidth]{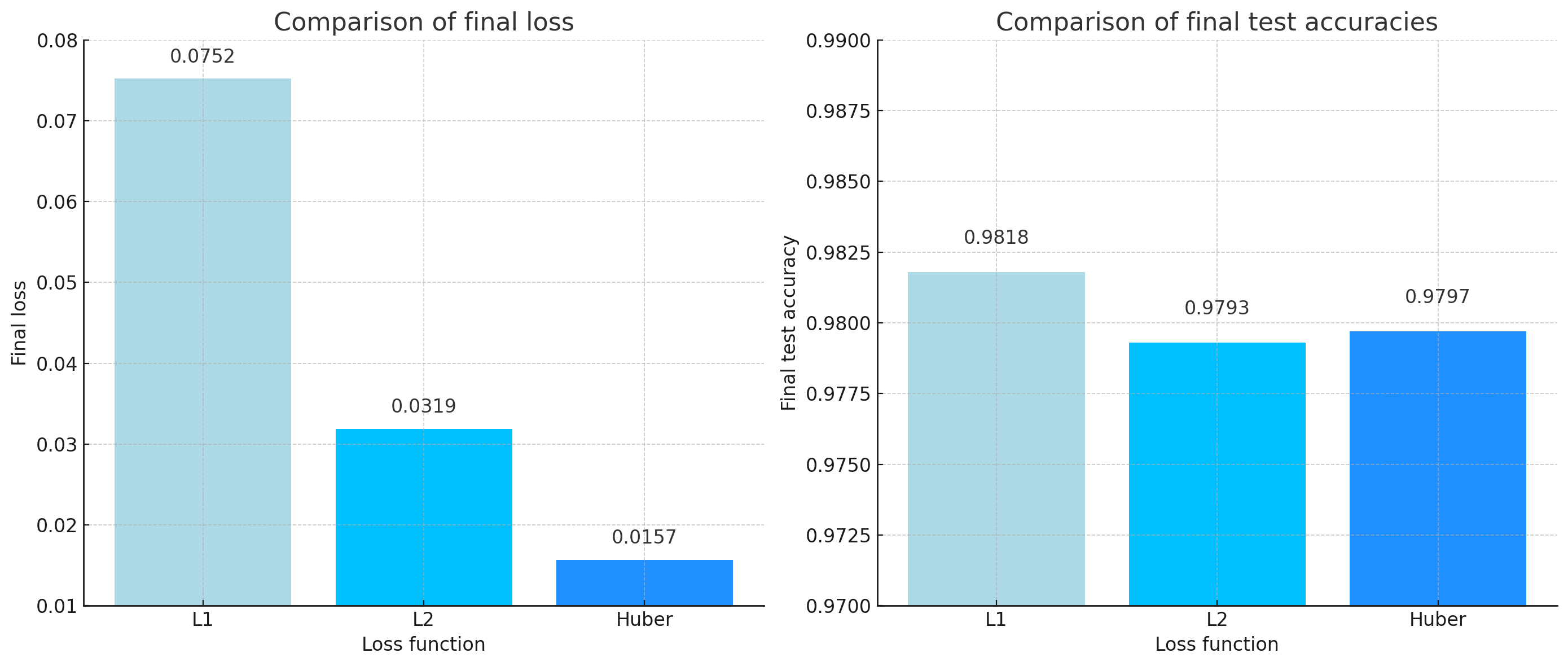}
\caption{Comparison of the final loss and the final test accuracy with the three different loss function}
\label{fig:loss_test_acc_VS_loss_functions}
\end{figure}

While different loss functions may yield better minimization in terms of loss, they do not necessarily change the decision boundaries in quantum machine learning models \cite{aktar2024graph}. This explains why we observe similar test accuracy across all the loss functions despite different loss values. The stability in test accuracy is also tied to the expressivity and generalization capabilities of quantum classifiers. The decision boundaries, particularly in data re-uploading quantum classifiers, remain stable across different loss functions, provided the quantum circuit architecture is the same, as the data's separability isn't significantly affected by the choice of loss function. This explains why, despite Huber loss achieving lower final loss values, the test accuracy remains close to that of L1 and L2 losses. The choice of loss function in quantum models often optimizes gradients but does not drastically alter the classification performance.\\

We do not use cross-entropy as the loss function here because our single-qubit QRU produces only one expectation value in $[-1,1]$, rather than a probability vector over the three target classes. To apply cross-entropy directly, the quantum circuit would need to output at least three class probabilities, which in practice requires adding more qubits (e.g., two qubits already yield four basis states, enough to encode three classes). This, however, would break comparability with our one-qubit setting and alter the reference pipeline. Exploring multi-qubit QRUs with cross-entropy will be the subject of future work.\\

Next, we turn to the performance of the QRU when using different learning rates.

\paragraph{Learning Rate} \hfill

\vspace{0.5cm}

\begin{figure}[H]
    \centering
    \begin{subfigure}[b]{1.0\textwidth}
        \centering
        \includegraphics[width=\textwidth]{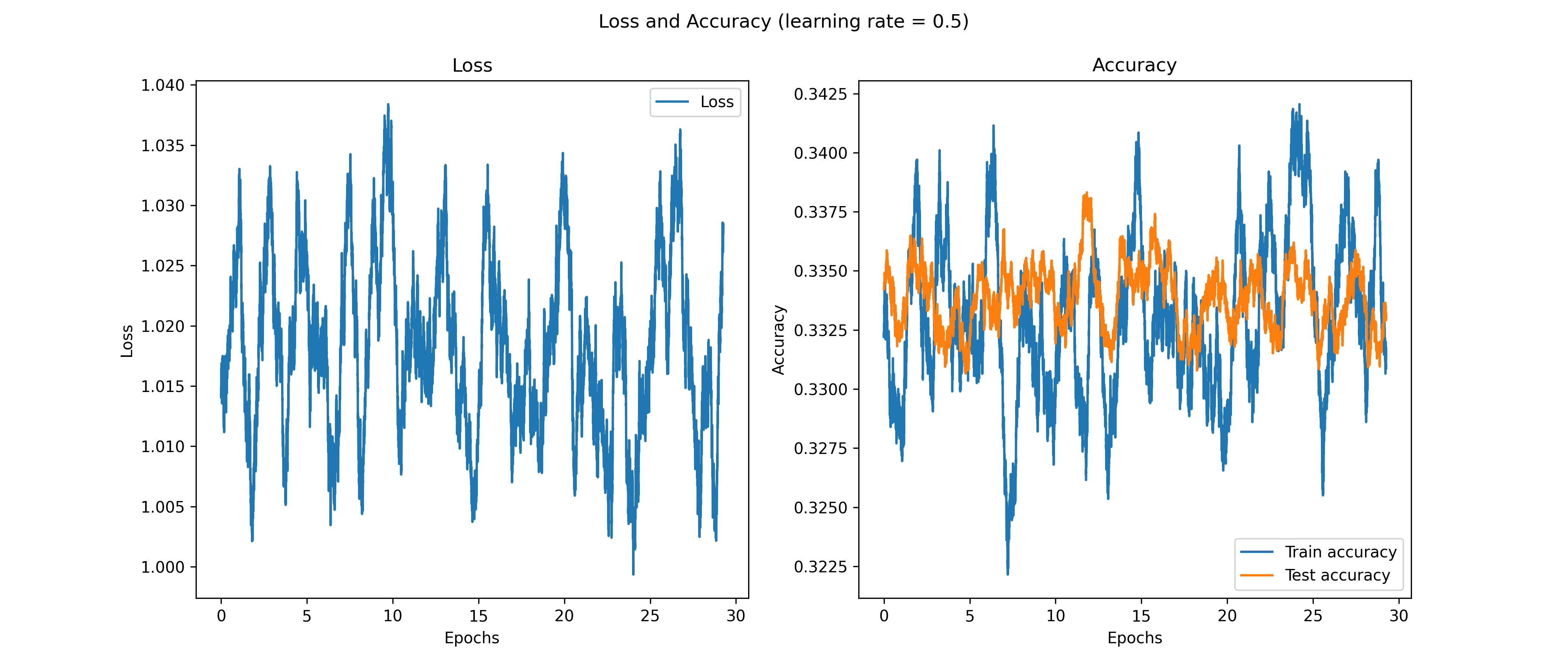}
        \caption{Learning rate = 0.5}
        \label{fig:lr_0.5}
    \end{subfigure}

    \begin{subfigure}[b]{1.0\textwidth}
        \centering
        \includegraphics[width=\textwidth]{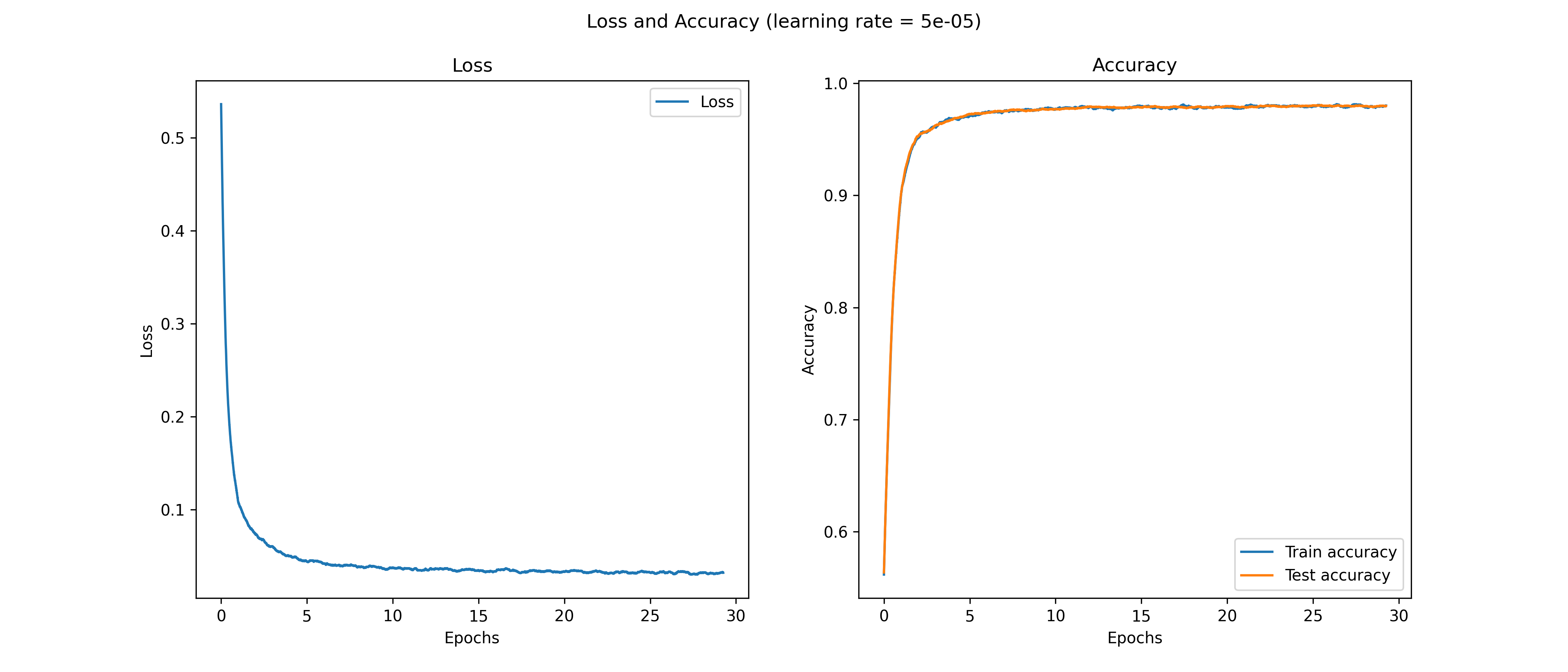}
        \caption{Learning rate = 5e-05}
        \label{fig:lr_5e-05}
    \end{subfigure}

    \begin{subfigure}[b]{1.0\textwidth}
        \centering
        \includegraphics[width=\textwidth]{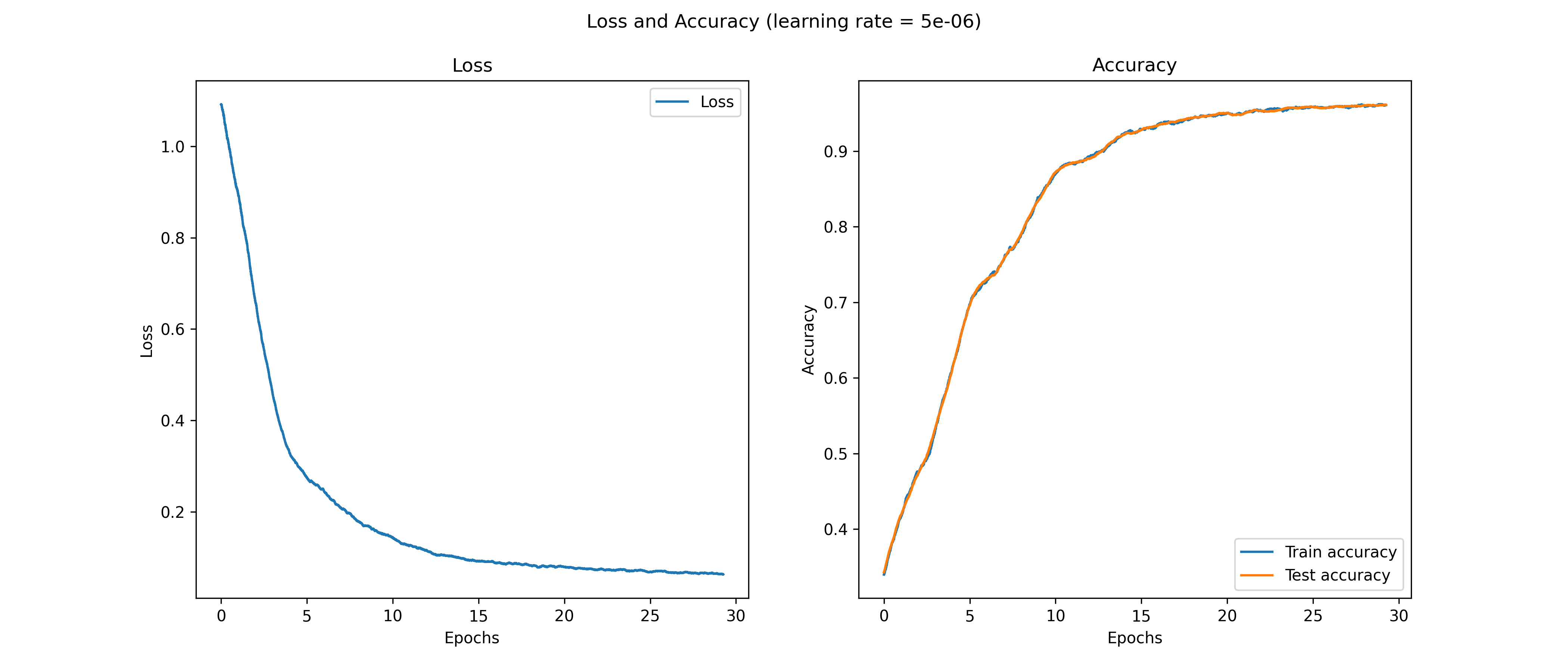}
        \caption{Learning rate = 5e-06}
        \label{fig:lr_5e-06}
    \end{subfigure}

    \caption{Loss and accuracy for different learning rates: 0.5 (\ref{fig:lr_0.5}), 5e-05 (\ref{fig:lr_5e-05})and 5e-06 (\ref{fig:lr_5e-06}).}
    \label{fig:lr_comparison}
\end{figure}

The goal of this section is to examine the impact of different initial learning rates on the model’s convergence, final accuracy and loss. The results shown in Figure \ref{fig:lr_comparison} highlight the impact of different learning rates on model performance. 

For a learning rate of 0.5 using the Adam optimizer (Figure \ref{fig:lr_0.5}), we observe significant oscillations in both loss and accuracy, indicating instability due to large parameter updates, with accuracy remaining around 0.33. In contrast, a learning rate of 5e-05 (Figure \ref{fig:lr_5e-05}) with Adam yields smooth convergence, reaching nearly 1.0 accuracy and showing a good balance between stability and learning speed. Finally, the smaller learning rate of 5e-06 (Figure \ref{fig:lr_5e-06}) also provides stable learning but slower convergence, with accuracy improving more gradually. These results suggest that 5e-05 is optimal for Adam, offering a good trade-off between speed and stability. \cite{dozat2016incorporating}\

\begin{figure}[H]
    \centering
    \begin{subfigure}[b]{0.45\textwidth}
        \centering
        \includegraphics[width=\textwidth]{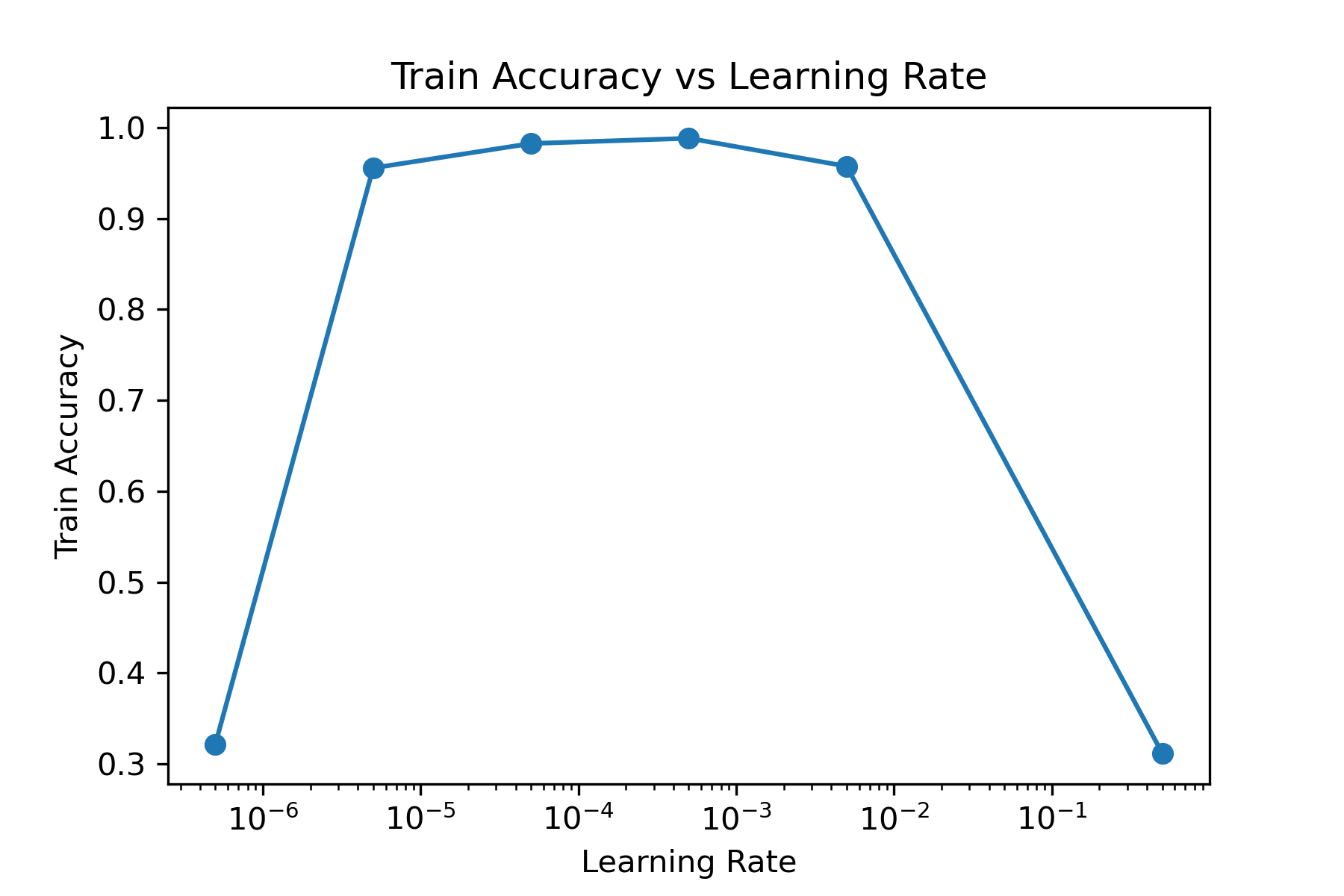}
        \caption{Accuracy evolution as a function of lr.}
        \label{fig:train_accuracy_lr}
    \end{subfigure}    
    \begin{subfigure}[b]{0.45\textwidth}
        \centering
        \includegraphics[width=\textwidth]{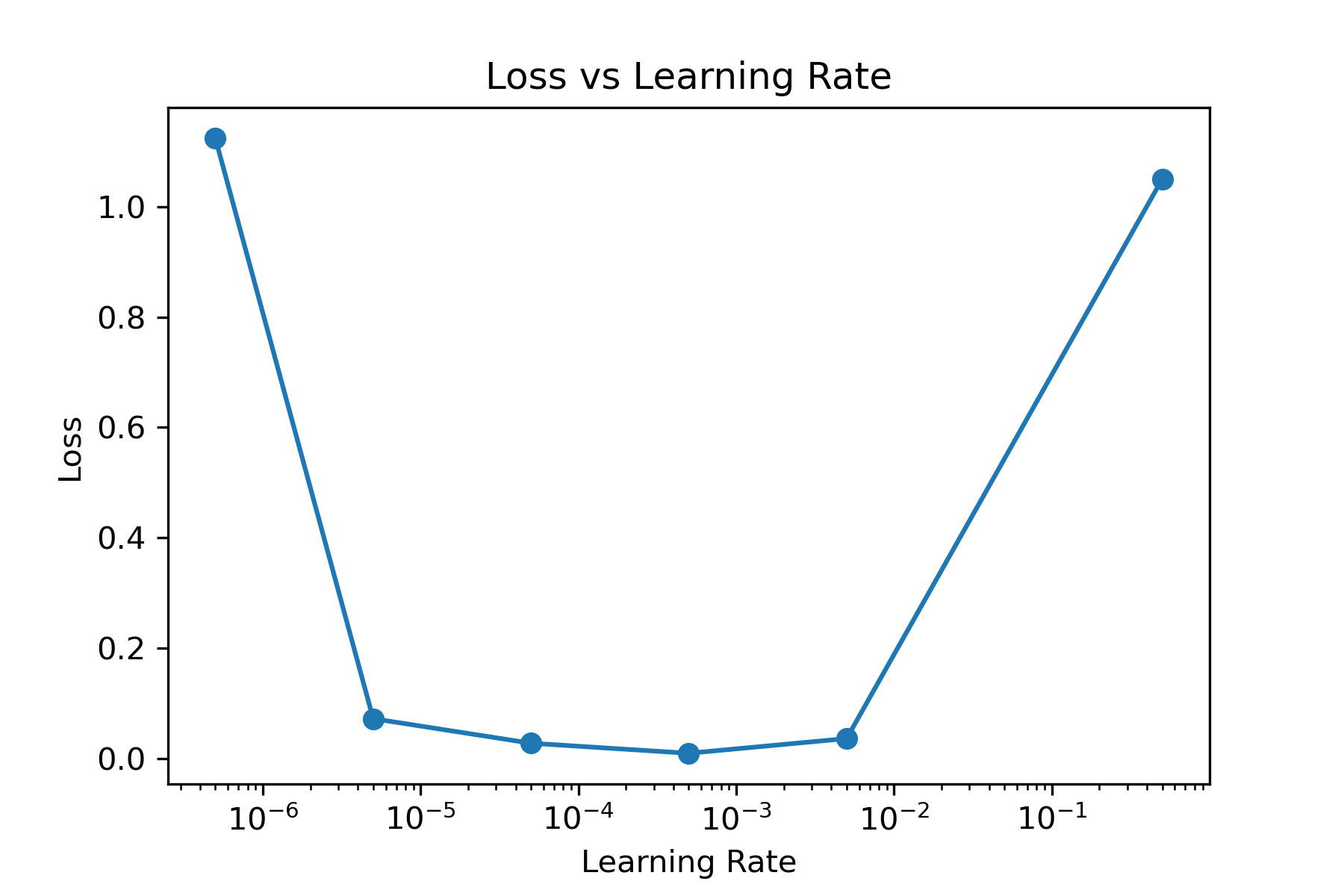}
        \caption{Loss evolution as a function of lr.}
        \label{fig:loss_lr}
    \end{subfigure}
    \caption{Comparison of accuracy (\ref{fig:train_accuracy_lr}) and loss (\ref{fig:loss_lr}) as a function of lr}
    \label{fig:comparison_lr}
\end{figure}

As shown in the graphs (Figures \ref{fig:train_accuracy_lr} and \ref{fig:loss_lr}), high learning rates such as \(5 \times 10^{-1}\) result in higher loss and instability in the model’s performance. In contrast, moderate learning rates like \(5 \times 10^{-5}\) and \(5 \times 10^{-6}\) show better convergence, with lower final loss and higher accuracy.

Table \ref{tab:lr} presents the final loss and accuracy metrics for different learning rates:

\begin{table}[h]
\centering
\caption{Final loss, training and test accuracies for different learning rates. Best values are in bold.}
\label{tab:lr}
\setlength{\tabcolsep}{5pt}
\begin{tabular}{lcccccc}
\toprule
Learning rate & $5{\times}10^{-1}$ & $5{\times}10^{-3}$ & $5{\times}10^{-4}$ & $5{\times}10^{-5}$ & $5{\times}10^{-6}$ & $5{\times}10^{-7}$ \\
\midrule
Loss            & 1.050 & 0.0358 & \textbf{0.0092} & 0.0274 & 0.0718 & 1.124 \\
Train accuracy  & 0.312 & 0.958 & \textbf{0.988} & 0.983 & 0.956 & 0.322 \\
Test accuracy   & 0.323 & 0.953 & \textbf{0.985} & 0.975 & 0.960 & 0.364 \\
\bottomrule
\end{tabular}
\end{table}

The experiments demonstrate that moderate learning rates yield the best performance. With a learning rate of \(5 \times 10^{-4}\), the model reaches a test accuracy of 98.5$\%$, outperforming both higher and lower rates. The results also show that for very small initial learning rates, there is a degradation in performance. This could be due to the fact that the model has not fully converged within the 30 epochs, or it might be trapped in a local minimum. An ongoing study is looking at whether launching the experiment with a small initial learning rate and different random seeds might yield a lower final loss compared to using higher initial learning rates.\\

In our experiments, we implemented an adaptive learning rate schedule to improve model convergence. The goal of this approach is to start with a relatively high learning rate, allowing the model to learn quickly at the beginning and then reduce the learning rate at specific milestones. This reduction helps the model fine-tune its parameters as it approaches convergence, leading to more stable results.

\begin{figure}[H]
\centering
\includegraphics[width=0.4\textwidth]{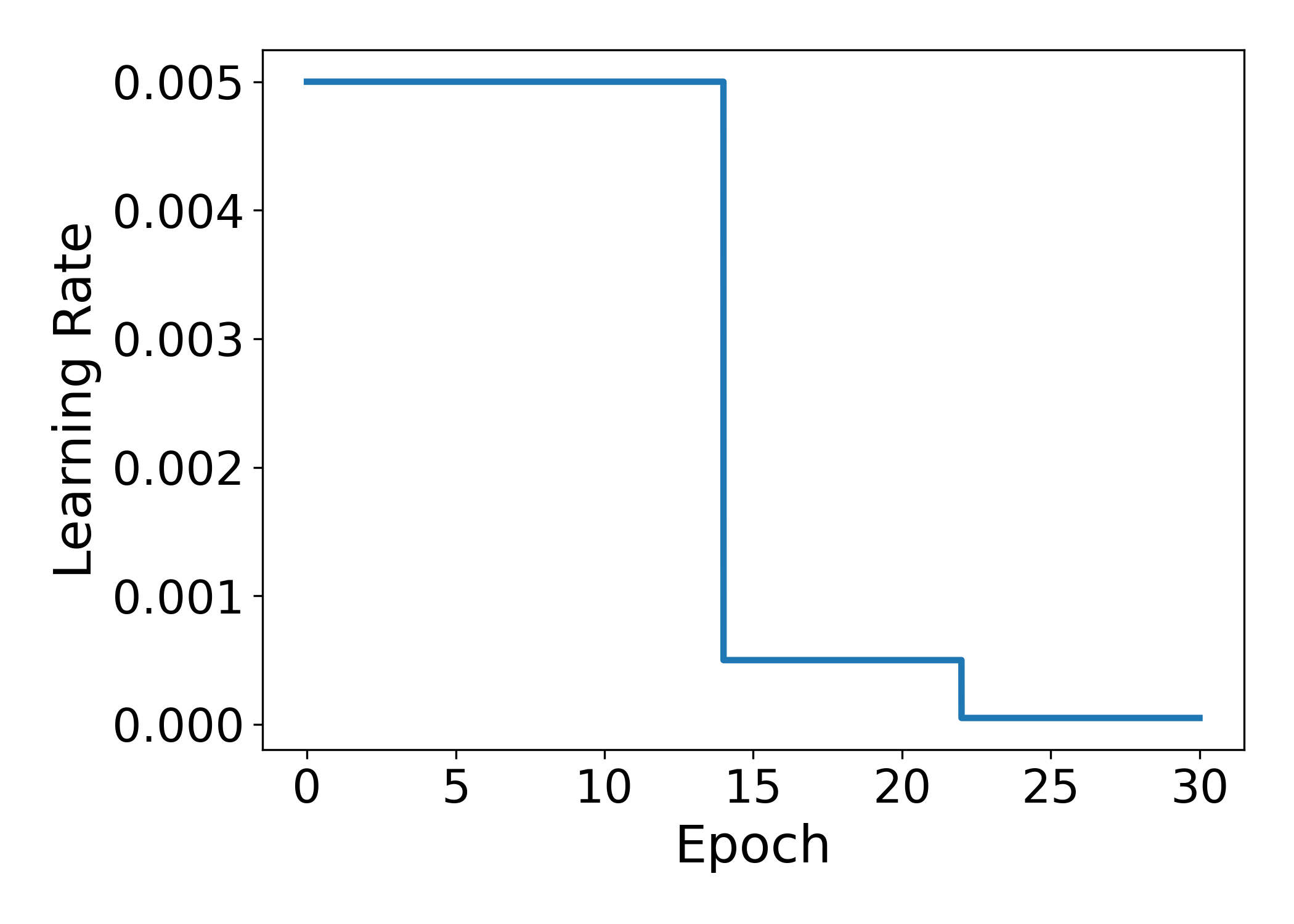}
\caption{Adaptive learning rate schedule showing the gradual reduction in learning rate over training epochs.}
\label{fig:adaptive_lr_schedule}
\end{figure}

The learning rate schedule is as follows:
\begin{enumerate}
    \item The initial learning rate is set to 0.005.
    \item At the halfway point in the total number of epochs, the learning rate is divided by 10 to slow down updates and enable more precise learning.
    \item At three-quarters of the way through the epochs, the learning rate is further divided by 10 to refine the convergence even more.
\end{enumerate}

figure \ref{fig:adaptive_lr_schedule} illustrates how the learning rate decreases across epochs, using the described stepwise reduction scheme.

The results we obtained show that manually changing the learning rate with the Adam optimizer does not significantly affect performance, since Adam adapts the effective step size for each parameter based on the magnitude and variance of past gradients. Therefore, we repeated the same experiment using SGD to investigate whether a scheduled learning rate could help. However, as already observed in Table~\ref{tab:optimizer_performance}, the SGD optimizer was not well suited to our task and did not perform adequately. \

We can still observe that progressively decreasing the learning rate during training led to slightly better results than using a fixed learning rate for SGD. Nevertheless, it remains an inefficient optimizer for our specific problem. 

\begin{figure}[H]
\centering
\includegraphics[width=0.8\textwidth]{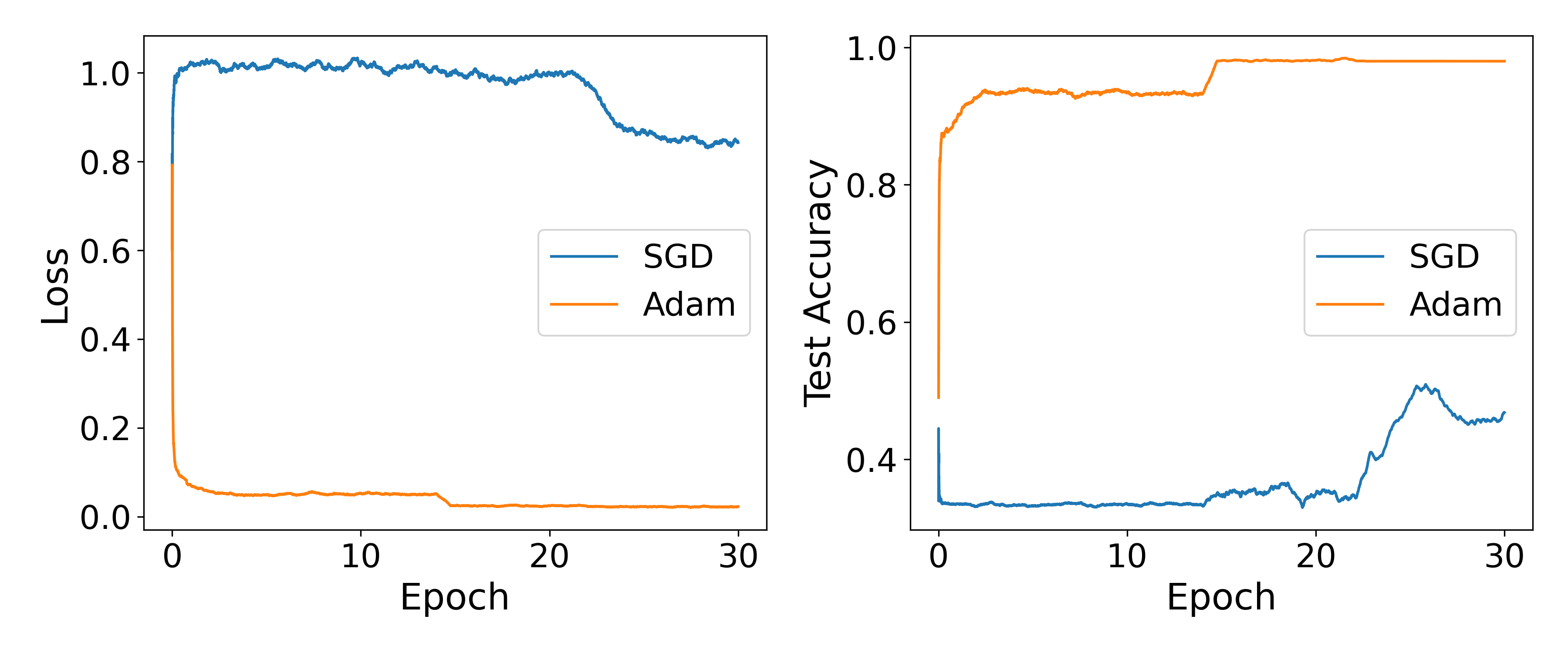}
\caption{Evolution of the loss and test accuracy for SGD and Adam optimizers with adaptive learning rate}
\label{fig:train_accuracy_lr_adaptative}
\end{figure}

In summary, this experiment clearly demonstrated that SGD was not effective in any configuration, even with a learning rate schedule, while Adam did not require any learning rate scheduling to perform as expected. Indeed, we obtained the same final test accuracy and loss values as without the scheduler, although the learning dynamics followed a slightly different descent trajectory.\\

We now turn to the overall optimization process to evaluate its impact on convergence and accuracy.

\subsection{Global optimization}

In this subsection, we discuss the application of global optimization methods to the QRU model to enhance hyperparameter tuning. We focus on Bayesian optimization \cite{snoek2012practical} as a primary method and briefly discuss our initial tests with Hyperband. We also analyze the observed correlations between key hyperparameters and their combined effects on model performance.

\subsubsection{Bayesian optimization}

Bayesian optimization \cite{bayesian_opt} is an efficient method for hyperparameter optimization when model evaluations are expensive. The objective function $f(x)$, which we seek to optimize, is modeled using a Gaussian process (GP). The GP is defined by a mean function $\mu(x)$ and a covariance function $k(x, x')$:

\begin{equation}
f(x) \sim \mathcal{GP}(\mu(x), k(x, x'))
\end{equation}

The Upper Confidence Bound (UCB) \cite{srinivas2009gaussian} acquisition function guides the selection of the next point to evaluate by maximizing the sum of the predicted mean $\mu(x)$ and an upper confidence bound proportional to the uncertainty $\sigma(x)$:

\begin{equation}
UCB(x) = \mu(x) + \kappa \sigma(x)
\end{equation}

Where $\kappa$ is a parameter balancing exploration (regions where $\sigma(x)$ is high) and exploitation (regions where $\mu(x)$ is high). Once a point $x_t$ is evaluated, the GP is updated using the newly observed data. The updated mean $\mu_{t+1}(x)$ is calculated by conditioning the GP on the previous points $X_t = \{x_1, x_2, \dots, x_t\}$ and their corresponding observations $y_t = \{f(x_1), f(x_2), \dots, f(x_t)\}$:

\begin{equation}
\mu_{t+1}(x) = k(x, X_t)^\top K(X_t, X_t)^{-1} y_t
\end{equation}

Here, $K(X_t, X_t)$ is the covariance matrix of the previously evaluated points and $k(x, X_t)$ is the covariance vector between the new point $x$ and the previously evaluated points $X_t$. This equation leverages the correlations between $x$ and past points to compute the new expected mean. The variance $\sigma_{t+1}^2(x)$ is updated similarly, representing the uncertainty of the GP's prediction at point $x$ after conditioning on the previously observed data:

\begin{equation}
\sigma_{t+1}^2(x) = k(x, x) - k(x, X_t)^\top K(X_t, X_t)^{-1} k(x, X_t)
\end{equation}

In this equation, $k(x, x)$ represents the variance at point $x$ (before conditioning) and the second term adjusts this variance based on how correlated $x$ is with the previously evaluated points. The result gives the updated uncertainty for $x$. This process is repeated until convergence, efficiently finding the optimal hyperparameters.\\

In our work, we have integrated Bayesian optimization into the training process to search for the best hyperparameters for our model. We defined the hyperparameter search space as follows:
\begin{itemize}
    \item \textbf{depth}: \{1, 2, 3, 4, 5, 6, 7, 8, 9, 10\}
    \item \textbf{learning rate}: \{0.5, 0.05, 0.005, 0.0005, 0.00005, 0.000005, 0.0000005\}
    \item \textbf{loss function}: \{L1, L2, Huber\}
    \item \textbf{optimizer}: \{SGD, RMSprop, Adam, Adamax, NAdam, Adagrad, Adadelta, AdamW\}
\end{itemize}

The objective function takes a set of hyperparameter values, trains the model with these values and returns the average loss. We used the \texttt{gp\_minimize} function from \texttt{scikit-optimize} to execute Bayesian optimization and find the best hyperparameters: \texttt{gp\_minimize(objective, space, n\_calls=50, random\_state=0, acq\_func="UCB", kappa=4)}.\\

\begin{figure}[H]
    \centering
    \includegraphics[width=0.95\textwidth]{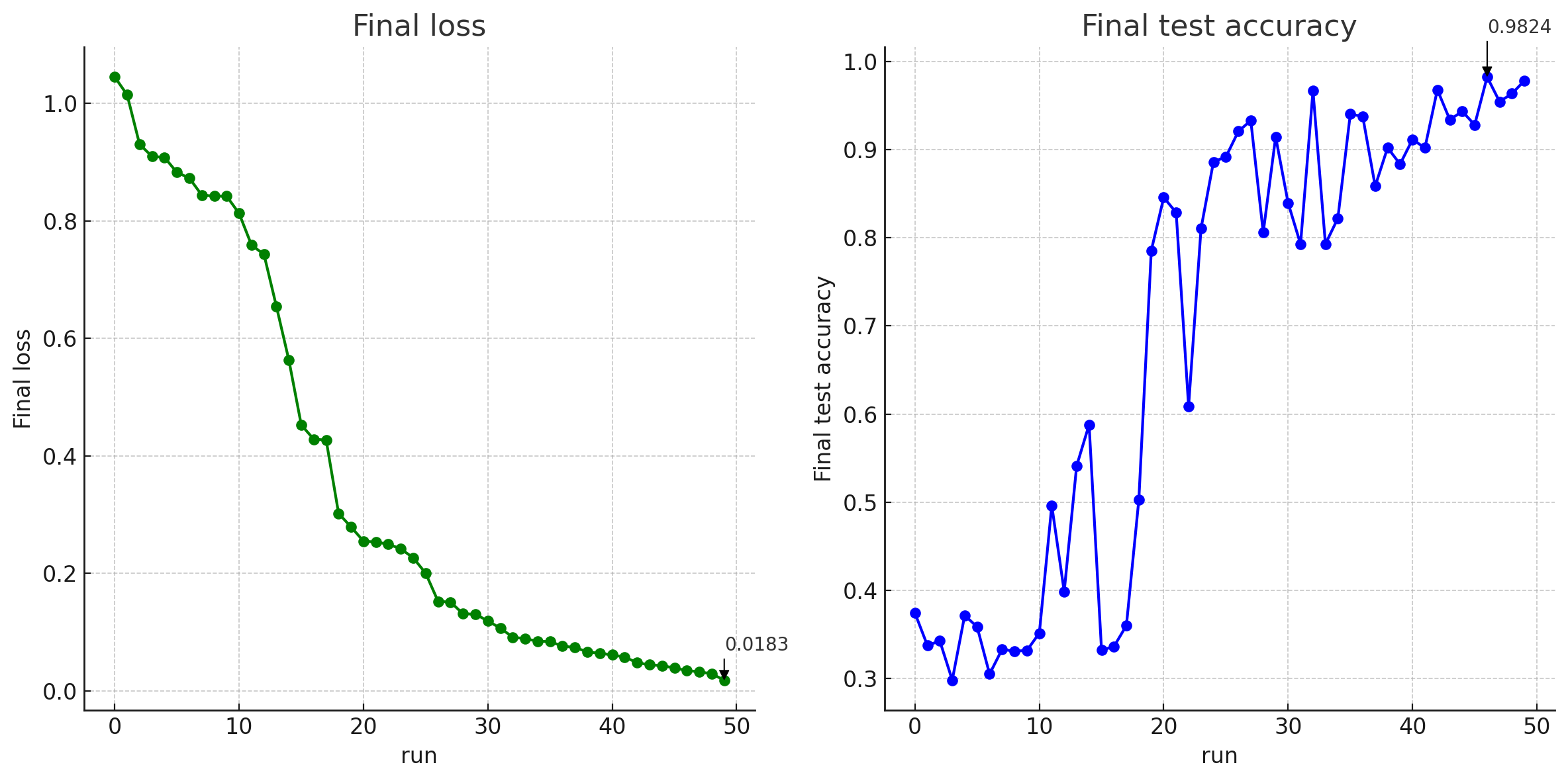}
    \caption{Evolution of the loss (left) and accuracy (right) during Bayesian optimization}
    \label{fig:convergence_plot}
\end{figure}

Figure~\ref{fig:convergence_plot} shows the obtained results: on the left, the evolution of the final loss, and on the right, the evolution of the final test accuracy as a function of different optimization runs.

The optimal configuration that resulted in the smallest final loss is: \texttt{depth=7, learning\_rate=0.5, loss\_function=Huber, optimizer=Adagrad} with a score of $0.018$. Although this configuration minimized the loss, the final accuracy associated with this configuration remains slightly lower than that of other configurations, reaching $0.978$. In fact, the best accuracy achieved was $0.9824$, obtained with the configuration \texttt{depth=5, learning\_rate=0.5, loss\_function=L2, optimizer=Adadelta}, but this configuration produced a slightly higher loss : $0.0344$.\\

One notable aspect of the Bayesian optimization results is the relatively high learning rate of $0.5$. Such a rate might seem too large in a classical optimization context. However, the use of the Adagrad and Adadelta optimizer compensates for this effect. Adagrad dynamically adjusts the learning rate based on the accumulated gradients during training. This mechanism allows the model to converge quickly at the beginning of training by exploring promising regions of the parameter space. This could be likely why Bayesian optimization identified a configuration with a high initial learning rate as optimal.\\

These results show that manual optimization can also lead to highly performant configurations, notably with an accuracy slightly higher than that obtained through Bayesian optimization, reaching $0.9854$ with a learning rate of $5e^{-4}$. Manual optimization thus provides a better understanding of the individual impact of each hyperparameter, while global optimization allowed us to see which combinations of hyperparameter values were correlated.\\

In conclusion, although global Bayesian optimization effectively reduced the loss and quickly explored the hyperparameter space, manual hyperparameter optimization allowed for slightly better accuracy maximization. These results demonstrate the importance of understanding hyperparameter impact and adopting a combined approach, using both Bayesian optimization to efficiently explore hyperparameter combinations and manual optimization to fine-tune the most sensitive parameter settings.

\subsubsection{Hyperband optimization} 

We also tested a global hyperparameter optimization method presented by Charles Moussa et al. for quantum neural networks \cite{Moussa_2022}. HyperBand stands out for its adaptive approach, allocating a computational budget to several hyperparameter configurations and progressively eliminating the less promising ones. Optimization can be further improved by integrating techniques such as fANOVA, priors and surrogates. \cite{hutter2014efficient}\\ 

To analyse the importance of hyperparameters and their interaction, we decomposed the variance of a model performance using an fANOVA model (Functional Analysis of Variance). It portions out this variance for each hyperparameter where the portion corresponds to their marginal effects, i.e., average effect of a given hyperparameter on performance. Priors on previous experiments or empirical knowledge help HyperBand to inform about hyperparameter selection better. While HyperBand with priors samples hyperparameters at random over the search space, it makes use of effective parameter values from other studies and emphasizes portions of the search space, which tends to lead to improved performing configurations. \ 

For instance, instead of sampling the learning rate uniformly at random within a range, HyperBand can restrict sampling to a prior distribution derived from the best values observed in previous trials. 
A surrogate model then approximates the objective function, predicting the expected performance of a given hyperparameter configuration without retraining the full model, thereby reducing computational cost.

\begin{figure}[H]
    \centering
        \includegraphics[width=0.95\textwidth]{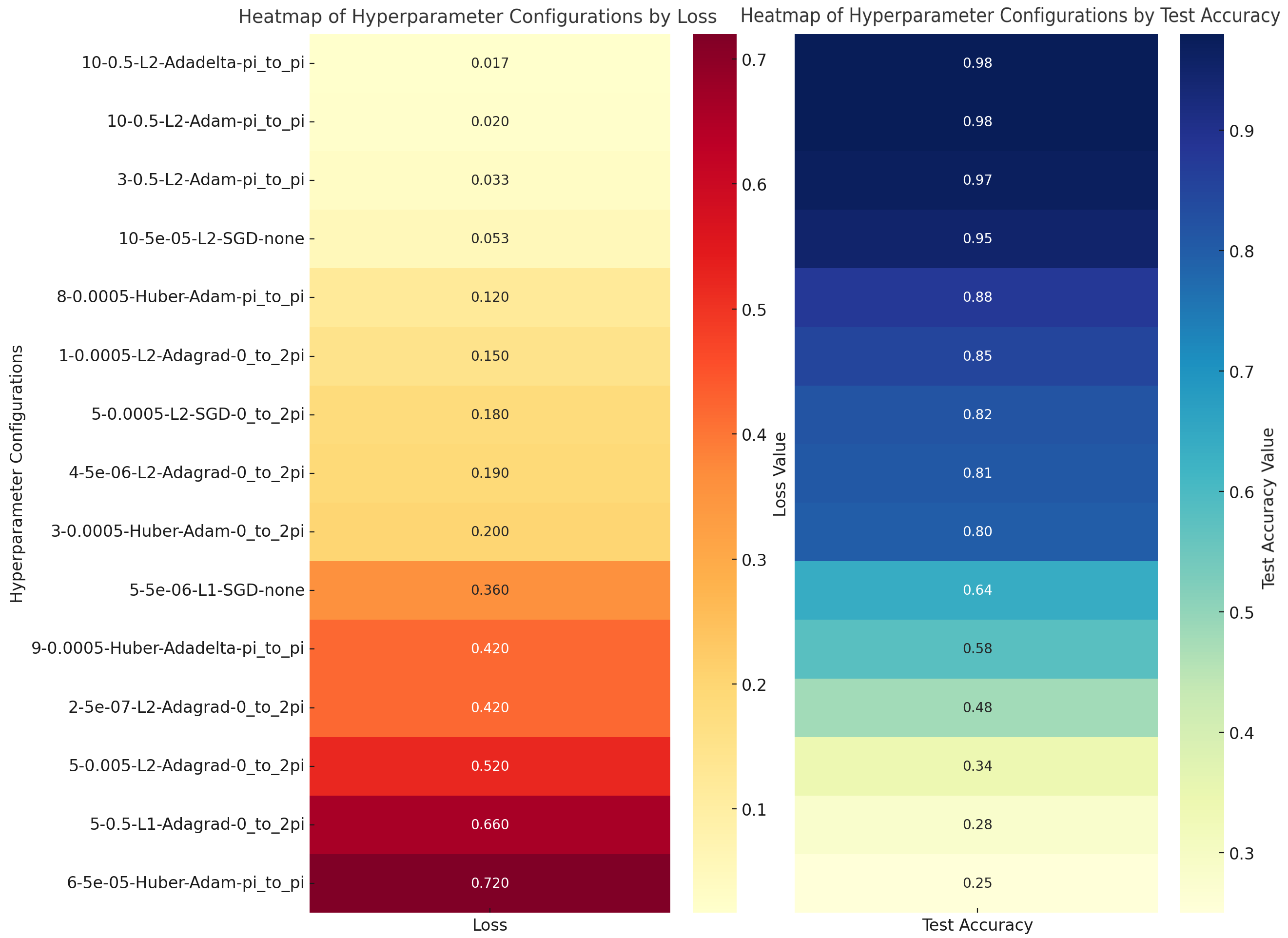}
    \caption{HyperBand optimisation results}
    \label{hyp}
\end{figure}

HyperBand recovers trends consistent with our Bayesian search but under a different search space: unlike the Bayesian runs, HyperBand also tuned input normalization. Consequently, results are not strictly comparable across the two methods. In our QRU setting, HyperBand frequently selected large learning rates (e.g., $0.5$) together with adaptive optimizers (Adadelta/Adagrad/Adam), and often favored deeper circuits (see Fig.~\ref{hyp}). An fANOVA analysis suggests that, within the explored ranges, most of the performance variance is explained by the learning rate and the optimizer, whereas depth, loss function, and normalization contributed less---without implying they are negligible in general. Practically, HyperBand identified high-performing configurations faster than our Bayesian procedure.

\subsection{Comparison of QRU, VQC, and MLP}
\label{subsec:param_matched_qru_vqc_mlp}

\begin{figure}[h]
  \centering
  \includegraphics[width=\linewidth]{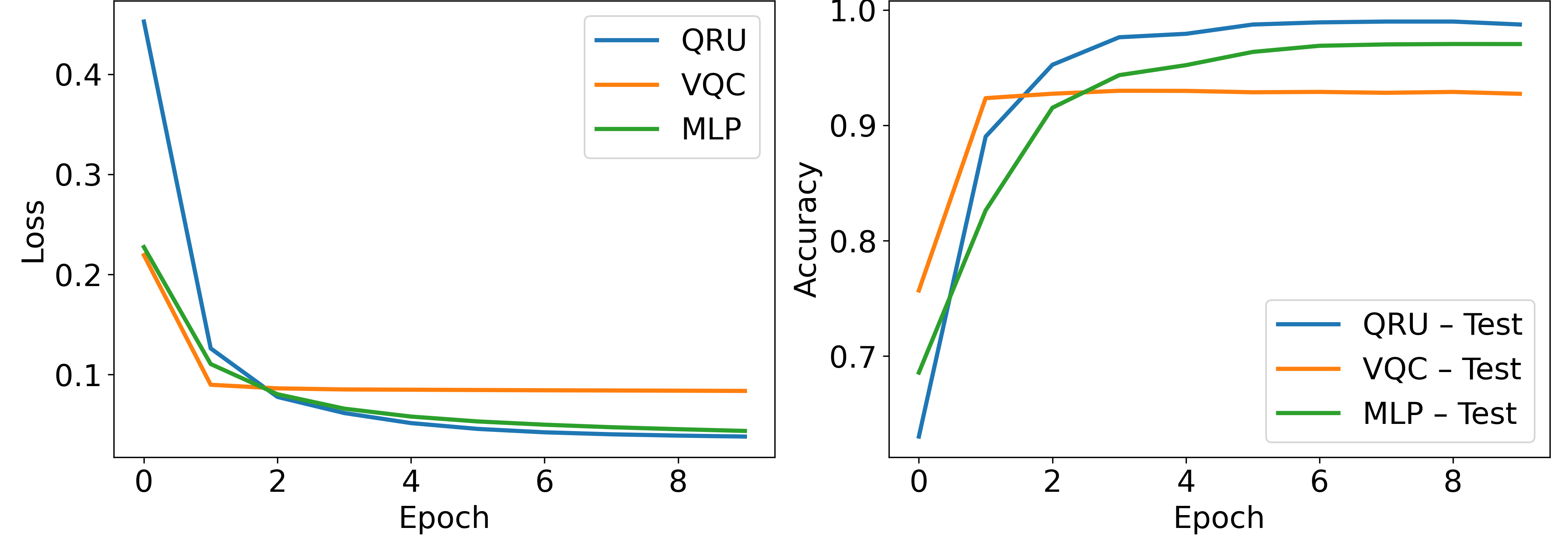}
  \vspace{-0.5em}
  \caption{Learning curves under matched parameter budgets.
  Left: loss vs.\ epoch.
  Right: accuracy vs.\ epoch}
  \label{fig:cmp_learning_curves}
\end{figure}

We compare three models under matched parameter budgets and an identical training protocol: a single-qubit QRU, a three-qubit VQC, and a classical MLP (3$\rightarrow h \rightarrow$1) whose trainable head is included in the parameter count. Figure~\ref{fig:cmp_learning_curves} reports per-epoch means of loss, train accuracy, and test accuracy.\\

\textbf{QRU (1 qubit):} Each re-uploading layer $i=1..L$ applies, for each feature $x_j$ ($j=1..D$), the sequence
$\mathrm{RX}(\theta^{(i)}_{j,0})$, $\mathrm{RY}(\theta^{(i)}_{j,1}\,x_j)$, $\mathrm{RX}(\theta^{(i)}_{j,2})$ on the single wire; readout is $\langle Z\rangle$. Trainable body parameters: $P_{\text{QRU}}=3DL$.\\

\textbf{VQC (3 qubits):} One-shot encoding $\mathrm{RY}(x_0)$ on $q_0$, $\mathrm{RY}(x_1)$ on $q_1$, $\mathrm{RY}(x_2)$ on $q_2$;
then $L$ variational layers of local trainable rotations
$\mathrm{RX}(\alpha_{i,q})$, $\mathrm{RY}(\beta_{i,q})$, $\mathrm{RZ}(\gamma_{i,q})$ on $q\in\{0,1,2\}$,
followed by ring entanglement (CNOT$_{0\to1}$, CNOT$_{1\to2}$, CNOT$_{2\to0}$). We aggregate onto $q_0$ via CNOT$_{1\to0}$ and CNOT$_{2\to0}$ and measure $\langle Z_0\rangle$. Trainable body parameters: $P_{\text{VQC}}=3QL$ with $Q{=}3$.\\

\textbf{MLP (3$\rightarrow h \rightarrow$1):} Two fully-connected layers with ReLU and a $\tanh$ output,
$\texttt{fc1}:\mathbb{R}^3\!\to\!\mathbb{R}^h$, $\texttt{fc2}:\mathbb{R}^h\!\to\!\mathbb{R}$, then $\tanh$. We include the head: $P_{\text{MLP}}=(3{+}1)h+(h{+}1)\cdot 1=5h{+}1$.\\

To match the body capacity of the two quantum models we set a common depth $L$. With $D{=}3$ and $Q{=}3$, this yields $P_{\text{QRU}}{=}3DL{=}9L$ and $P_{\text{VQC}}{=}3QL{=}9L$. We choose $h$ such that $5h{+}1\!\approx\!P_{\text{QRU}}$. Table~\ref{tab:param_budget} summarizes the counts used in our run ($L{=}10$).

\begin{table}[h]
\centering
\caption{Parameter-matched setup ($D{=}3$, $Q{=}3$, $L{=}10$). The MLP includes its head in $P_{\text{MLP}}$.}
\label{tab:param_budget}
\setlength{\tabcolsep}{3pt}
\begin{tabular}{@{}lcccc@{}}
\toprule
Model & Formula & Hyperparams & $P$ & Output \\
\midrule
QRU (1q) & $3DL$ & $D{=}3,\;L{=}10$ & $3\!\times\!3\!\times\!10=90$ & $\langle Z\rangle$ \\
VQC (3q) & $3QL$ & $Q{=}3,\;L{=}10$ & $3\!\times\!3\!\times\!10=90$ & $\langle Z_0\rangle$ \\
MLP (3$\!\to\!h\!\to\!1$) & $5h{+}1$ & $h{=}\lfloor(90{-}1)/5\rfloor{=}17$ & $5\!\times\!17{+}1=86$ & $\tanh$ \\
\bottomrule
\end{tabular}
\end{table}

\noindent Under this protocol (LR $=5{\times}10^{-5}$, $10$ epochs), we obtain results shown with Table~\ref{tab:cmp_results}:
\begin{table}[h]
\centering
\caption{Final per-epoch means under matched parameter budgets.}
\label{tab:cmp_results}
\setlength{\tabcolsep}{5pt}
\begin{tabular}{@{}lcccc@{}}
\toprule
Model& Loss & Train Acc & Test Acc \\
\midrule
QRU (1q) & \textbf{0.038} & \textbf{0.977} & \textbf{0.987} \\
VQC (3q)  & 0.084 & 0.956 & 0.927 \\
MLP ($h{=}17$)  & 0.044 & 0.948 & 0.970 \\
\bottomrule
\end{tabular}
\end{table}

A coarse trainability proxy (area under the loss curve across epochs) ranks MLP~$<$~QRU~$\lesssim$~VQC, i.e., the MLP optimizes fastest (lowest loss area), while the QRU achieves the best generalization in this run.\

QRU attains the highest test accuracy (0.987), followed by MLP (0.970). VQC lags (0.927) and shows a negative generalization gap (test $<$ train, $-0.029$), unlike QRU and MLP where test is slightly higher than train. \

MLP reaches the lowest loss and the smallest area-under-loss, consistent with faster optimization on this scalar-regression-with-thresholds objective, however, low loss does not translate into the best test accuracy here.\

With matched body budgets ($P{=}90$) and identical logging, re-uploading on a single qubit (QRU) proves highly competitive, slightly outperforming the 3-qubit VQC with ring entanglement under L2+threshold training.

\subsection{Hardware inference on a real QPU (qBraid-managed IQM Garnet)}
\label{subsec:qpu_inference_iqm}

This experiment validates that the trained QRU classifier can be executed on real quantum hardware while preserving the exact circuit structure and decision rule used in simulation.\\

We train the 1-qubit QRU classifier using PennyLane's \texttt{default.qubit} simulator. The circuit implements a data re-uploading scheme with input dimension $D=3$ and depth $L=7$. For each layer $d\in\{1,\dots,L\}$ and each feature $j\in\{1,\dots,D\}$, the model applies the gate sequence
\[
\mathrm{RX}(\theta_{d,0,j}) \;\rightarrow\; \mathrm{RY}(x_j\,\theta_{d,1,j}) \;\rightarrow\; \mathrm{RX}(\theta_{d,2,j}),
\]
followed by a measurement of $\langle Z\rangle$ on the single qubit. This matches the training implementation exactly. \texttt{Adadelta} optimization is used with an $L_2$ loss, and classification is obtained by thresholding the scalar output $\hat{y}\in[-1,1]$ into labels $\{-1,0,+1\}$ using the fixed rule: $\hat{y}<-0.33\mapsto -1$, $\hat{y}>0.33\mapsto +1$, and $|\hat{y}|\le 0.33\mapsto 0$.%
\footnote{See the training implementation for the circuit definition and the label mapping.}
After training, we export the best parameters $\theta$ along with the model configuration (depth, input size) into a portable payload for hardware execution (PyTorch state + JSON export).\\

To run inference on a real device while keeping the same trained parameters, we:
(i) rebuild the circuit with the frozen parameters $\theta$ and each test input $x$,
(ii) export the resulting quantum tape to \texttt{OpenQASM 2.0},
(iii) submit the batch of QASM circuits through \texttt{qbraid.runtime} to the qBraid-managed device \texttt{iqm\_garnet},
and (iv) reconstruct the expectation value from measurement counts using
\[
\widehat{\langle Z\rangle} \;=\; \hat{y} \;=\; \frac{c_0 - c_1}{N_{\mathrm{shots}}},
\]
where $c_0$ and $c_1$ are the observed counts for computational outcomes $0$ and $1$.
Finally, we apply the same thresholding rule as in CPU evaluation to obtain the predicted label.
This procedure ensures that any deviation between CPU and QPU inference originates from finite-shot noise and hardware noise, not from a change in model definition.\\

We evaluated the same first $N=10$ test samples (from the saved PyTorch test loader) on:
(i) CPU simulation (exact $\langle Z\rangle$), and
(ii) the IQM Garnet QPU via qBraid with $N_{\mathrm{shots}}=50$ and $200$.
All three runs yield $10/10$ correct predictions on this subset, indicating that (a) the trained parameters transfer correctly to hardware execution, and (b) the decision margins for these points are sufficiently far from the $\pm 0.33$ thresholds to remain stable under shot and device noise.\\

\begin{figure}[H]
    \centering
    \includegraphics[width=\linewidth]{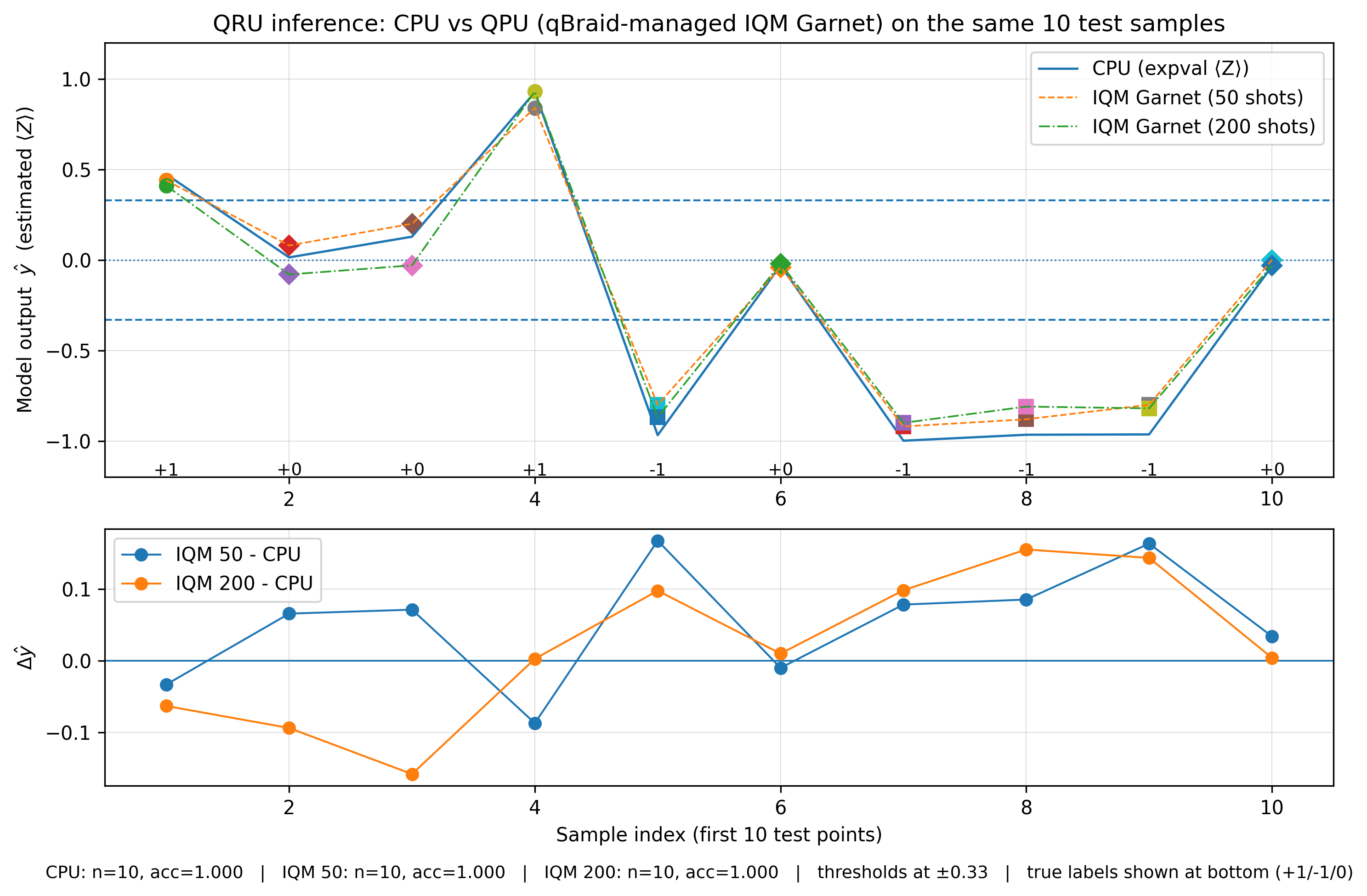}
    \caption{CPU vs QPU inference on the same 10 test samples using the trained QRU parameters ($L=7$, $D=3$). Top: model outputs $\hat{y}\approx \langle Z\rangle$ for CPU simulation and IQM Garnet runs (qBraid-managed) at 50 and 200 shots, with decision thresholds at $\pm 0.33$. Bottom: deviations $\Delta\hat{y}$ relative to CPU. True labels are indicated at the bottom of the top panel.}
    \label{fig:qpu_eval_iqm}
\end{figure}

Figure~\ref{fig:qpu_eval_iqm} compares the raw scalar outputs $\hat{y}$ across CPU and QPU, and reports the per-sample deviations $\Delta\hat{y}=\hat{y}_{\mathrm{QPU}}-\hat{y}_{\mathrm{CPU}}$. As expected, the QPU estimates fluctuate around the CPU reference, with visibly larger dispersion at $50$ shots than at $200$ shots. Importantly, these fluctuations do not cross the decision boundaries for the tested examples. While this is only a small-scale validation (and not a full hardware benchmarking campaign), it demonstrates end-to-end deployability of the learned QRU inference pipeline on real hardware.

\newpage
\section{Conclusion}
\label{sec:conclusion}

This work studied single-qubit quantum re-uploading units (QRUs) for a three-feature calorimetry classification task and compared them, at matched parameter count, to a standard mono-encoded variational quantum circuit (VQC) baseline. In this experimental setting, QRUs reached high accuracy with a small circuit footprint, and the ablation results indicate that most performance gains occur at small depths before saturating, while training cost grows approximately linearly with depth.

We also examined QRUs through a frequency-domain expressivity lens. Repeated data re-encoding expands the set of reachable trigonometric components relative to mono-encoding, and our spectral activation study is consistent with this mechanism. We stress, however, that this analysis concerns representational capacity and does not by itself address optimization difficulty, noise sensitivity, or generalization beyond the considered task.

Finally, we report an end-to-end proof-of-execution of the trained single-qubit QRU on a superconducting QPU via a cloud workflow. This hardware run is not intended as a noise-robust performance benchmark, but it demonstrates practical deployability under current constraints.

Key limitations include the use of a small set of engineered features and a scalar-to-multiclass thresholding scheme, as well as a restricted baseline class focused on mono-encoding. Future work should extend the evaluation to multi-logit readouts, less curated inputs, noise-aware training and mitigation, multi-dimensional spectral probes, and broader baselines including re-encoded multi-qubit VQCs and capacity-matched classical models.

\section*{Statements and Declarations}

\noindent \textbf{Data availability:} This work was supported by the TAIAO programme and by the PhD scholarship funding provided through the TAIAO project.

\noindent \textbf{Competing interests:} The authors have no financial or non-financial competing interests to declare that are relevant to the content of this article.

\noindent \textbf{Data availability:} All datasets analysed in this study are openly available: \url{https://llraidata.in2p3.fr/hgcnn/d2/}.

\noindent \textbf{Code availability:} All code used for data processing, model training, and figure generation is available at the same repository:  
\url{https://github.com/LeaCasse/QRU_Calorimetry_Optimization}.

\noindent\textbf{Author contributions:} All authors contributed to the the study.  

\noindent \textbf{Ethics approval and consent:} Not applicable.

\noindent \textbf{Acknowledgements and Funding Declaration:} The authors would like to thank the TAIAO programme and the University of Waikato for providing the research environment in which this work was conducted.

 \newpage
\scriptsize{\bibliography{references}}

@article{Biamonte2017,
   title={Quantum machine learning},
   volume={549},
   ISSN={1476-4687},
   url={http://dx.doi.org/10.1038/nature23474},
   DOI={10.1038/nature23474},
   number={7671},
   journal={Nature},
   publisher={Springer Science and Business Media LLC},
   author={Biamonte, Jacob and Wittek, Peter and Pancotti, Nicola and Rebentrost, Patrick and Wiebe, Nathan and Lloyd, Seth},
   year={2017},
   month=sep, pages={195–202} }

@misc{D2dset,
author = {Becheva, Emilia and Beaudette, Florian and Sauvan,
Jean-Baptiste and Melennec, Matthieu and Ghosh, Shamik and Mellin,
Michael and Magniette, Frederic},
title = {High granularity calorimetry D2 dataset},
doi = {10.5281/zenodo.14260279},
howpublished= {\url{https://zenodo.org/records/14260279}},
year = {2024}
}

@article{cerezo2021variational,
  title={Variational quantum algorithms},
  author={Cerezo, Marco and Arrasmith, Andrew and Babbush, Ryan and Benjamin, Simon C and Endo, Suguru and Fujii, Keisuke and McClean, Jarrod R and Mitarai, Kosuke and Yuan, Xiao and Cincio, Lukasz and others},
  journal={Nature Reviews Physics},
  volume={3},
  number={9},
  pages={625--644},
  year={2021},
  publisher={Nature Publishing Group UK London}
}

@techreport{HGCal,
       author = {CMS Collaboration},
       title         = {The Phase-2 Upgrade of the CMS Endcap
Calorimeter},
       institution   = {CERN},
       reportNumber  = {CERN-LHCC-2017-023, CMS-TDR-019},
       address       = {Geneva},
       year          = {2017},
       url           = {https://cds.cern.ch/record/2293646},
       doi           = {10.17181/CERN.IV8M.1JY2}
}

@misc{HGCalSimulations2021,
  author = {Project OGCID Team},
  title = {HGCal-like simulations},
  year = {2021},
  url = {https://llrogcid.in2p3.fr/hgcal-like-simulations/},
  note = {Accessed: 2024-09-23},
}

@misc{GNNParticleSim2023,
  author       = {Sanchez-Gonzalez, Alvaro and Battaglia, Peter and others},
  title        = {Simulating complex physics with graph networks},
  howpublished = {\url{https://github.com/deepmind/graph_nets}},
  note         = {DeepMind tutorial repository. Accessed: 2024-09-23},
  year         = {2020}
}

@misc{GNSMeshNet2023,
  author = {GeoElements Team},
  title = {Graph Network Simulator (GNS)},
  year = {2023},
  url = {https://github.com/geoelements/gns},
  note = {Accessed: 2024-09-23},
}

@article{briscoe2011conceptual,
  title={Conceptual complexity and the bias/variance tradeoff},
  author={Briscoe, Erica and Feldman, Jacob},
  journal={Cognition},
  volume={118},
  number={1},
  pages={2--16},
  year={2011},
  publisher={Elsevier}
}

@article{preskill18,
    title={Quantum Computing in the NISQ era and beyond},
    volume={2},
    ISSN={2521-327X},
    url={http://dx.doi.org/10.22331/q-2018-08-06-79},
    DOI={10.22331/q-2018-08-06-79},
    journal={Quantum},
    publisher={Verein zur Forderung des Open Access Publizierens in den
Quantenwissenschaften},
    author={Preskill, John},
    year={2018},
    month=aug, pages={79} }

@article{zhov2000quantum,
  title={Quantum neural networks},
  author={Ezhov, Alexandr A and Ventura, Dan},
  journal={Future Directions for Intelligent Systems and Information Sciences: The Future of Speech and Image Technologies, Brain Computers, WWW, and Bioinformatics},
  pages={213--235},
  year={2000},
  publisher={Springer}
}

@article{perez2019data,
  title={Data re-uploading for a universal quantum classifier},
  author={P{\'e}rez-Salinas, Adri{\'a}n and Cervera-Lierta, Alba and Gil-Fuster, Elies and Latorre, Jos{\'e} I},
  journal={Quantum},
  volume={3},
  pages={139},
  year={2019}
}

@misc{llr_cluster,
  author       = {{Laboratoire Leprince-Ringuet}},
  title        = {Computing Facilities at LLR},
  howpublished = {\url{https://llr.in2p3.fr/3-lab-computing}},
  note         = {Accessed: October 10, 2025},
  year         = {2025}
}

@inproceedings{chen2024quantum,
  title={Quantum-classical-quantum workflow in quantum-hpc middleware with gpu acceleration},
  author={Chen, Kuan-Cheng and Li, Xiaoren and Xu, Xiaotian and Wang, Yun-Yuan and Liu, Chen-Yu},
  booktitle={2024 International Conference on Quantum Communications, Networking, and Computing (QCNC)},
  pages={304--311},
  year={2024},
  organization={IEEE}
}

@article{asadi2024hybrid,
  title={Hybrid quantum programming with pennylane lightning on hpc platforms},
  author={Asadi, Ali and Dusko, Amintor and Park, Chae-Yeun and Michaud-Rioux, Vincent and Schoch, Isidor and Shu, Shuli and Vincent, Trevor and O'Riordan, Lee James},
  journal={arXiv preprint arXiv:2403.02512},
  year={2024}
}

@article{lobanov2020precision,
  title={Precision timing calorimetry with the CMS HGCAL},
  author={Lobanov, Artur},
  journal={Journal of Instrumentation},
  volume={15},
  number={07},
  pages={C07003},
  year={2020},
  publisher={IOP Publishing}
}

@inproceedings{hutter2014efficient,
  title={An efficient approach for assessing hyperparameter importance},
  author={Hutter, Frank and Hoos, Holger and Leyton-Brown, Kevin},
  booktitle={International conference on machine learning},
  pages={754--762},
  year={2014},
  organization={PMLR}
}

@inproceedings{casse2025quantum,
  title={Quantum Reupload Units: A Scalable and Expressive Approach for Time Series Learning},
  author={Cass{\'e}, L{\'e}a and Ponnambalam, Sabarikirishwaran and Pfahringer, Bernhard and Bifet, Albert},
  booktitle={2025 IEEE International Conference on Quantum Computing and Engineering (QCE)},
  volume={1},
  pages={1815--1825},
  year={2025},
  organization={IEEE}
}

@article{srinivas2009gaussian,
  title={Gaussian process optimization in the bandit setting: No regret and experimental design},
  author={Srinivas, Niranjan and Krause, Andreas and Kakade, Sham M and Seeger, Matthias},
  journal={arXiv preprint arXiv:0912.3995},
  year={2009}
}

@article{snoek2012practical,
  title={Practical bayesian optimization of machine learning algorithms},
  author={Snoek, Jasper and Larochelle, Hugo and Adams, Ryan P},
  journal={Advances in neural information processing systems},
  volume={25},
  year={2012}
}

@inproceedings{huber1992robust,
  title={Robust estimation of a location parameter},
  author={Huber, Peter J},
  booktitle={Breakthroughs in statistics: Methodology and distribution},
  pages={492--518},
  year={1992},
  organization={Springer}
}

@misc{dozat2016incorporating,
  author       = {Dozat, Timothy},
  title        = {Incorporating Nesterov Momentum into Adam},
  year         = {2016},
  howpublished = {ICLR Workshop Poster},
  url          = {https://openreview.net/forum?id=OM0jvwB8jIp57ZJjtNEZ}
}

@misc{loshchilov2019decoupledweightdecayregularization,
      title={Decoupled Weight Decay Regularization}, 
      author={Ilya Loshchilov and Frank Hutter},
      year={2019},
      eprint={1711.05101},
      archivePrefix={arXiv},
      primaryClass={cs.LG},
      url={https://arxiv.org/abs/1711.05101}, 
}

@article{geman1992neural,
  title={Neural networks and the bias/variance dilemma},
  author={Geman, Stuart and Bienenstock, Elie and Doursat, Ren{\'e}},
  journal={Neural computation},
  volume={4},
  number={1},
  pages={1--58},
  year={1992},
  publisher={MIT Press}
}

@article{wang2021noise,
  title={Noise-induced barren plateaus in variational quantum algorithms},
  author={Wang, Samson and Fontana, Enrico and Cerezo, Marco and Sharma, Kunal and Sone, Akira and Cincio, Lukasz and Coles, Patrick J},
  journal={Nature communications},
  volume={12},
  number={1},
  pages={6961},
  year={2021},
  publisher={Nature Publishing Group UK London}
}

@article{kingma2014adam,
  title={Adam: A method for stochastic optimization},
  author={Kingma, Diederik P},
  journal={arXiv preprint arXiv:1412.6980},
  year={2014}
}

@misc{glendinning2005bloch,
  author       = {Glendinning, Ian},
  title        = {The Bloch Sphere},
  year         = {2005},
  howpublished = {Lecture notes, Quantum Information and Algorithms Group},
  note         = {Unpublished manuscript}
}

@article{caro2021encoding,
  title={Encoding-dependent generalization bounds for parametrized quantum circuits},
  author={Caro, Matthias C and Gil-Fuster, Elies and Meyer, Johannes Jakob and Eisert, Jens and Sweke, Ryan},
  journal={Quantum},
  volume={5},
  pages={582},
  year={2021},
  publisher={Verein zur F{\"o}rderung des Open Access Publizierens in den Quantenwissenschaften}
}

@article{bharti2022noisy,
  title={Noisy intermediate-scale quantum algorithms},
  author={Bharti, Kishor and Cervera-Lierta, Alba and Kyaw, Thi Ha and Haug, Tobias and Alperin-Lea, Sumner and Anand, Abhinav and Degroote, Matthias and Heimonen, Hermanni and Kottmann, Jakob S and Menke, Tim and others},
  journal={Reviews of Modern Physics},
  volume={94},
  number={1},
  pages={015004},
  year={2022},
  publisher={APS}
}

@article{xu2024frequency,
  title={Frequency principle for quantum machine learning via Fourier analysis},
  author={Xu, Yi-Hang and Zhang, Dan-Bo},
  journal={arXiv preprint arXiv:2409.06682},
  year={2024}
}

@article{heimann2025learning,
  title={Learning Fourier series with parametrized quantum circuits},
  author={Heimann, Dirk and Hohenfeld, Hans and Sch{\"o}nhoff, Gunnar and Mounzer, Elie and Kirchner, Frank},
  journal={Physical Review Research},
  volume={7},
  number={2},
  pages={023151},
  year={2025},
  publisher={APS}
}

@article{schuld2021effect,
  title={Effect of data encoding on the expressive power of variational quantum-machine-learning models},
  author={Schuld, Maria and Sweke, Ryan and Meyer, Johannes Jakob},
  journal={Physical Review A},
  volume={103},
  number={3},
  pages={032430},
  year={2021},
  publisher={APS}
}

@article{zhou2021machine,
  title={Machine Learning},
  author={Zhou, Z.H. and Liu, S.},
  isbn={9789811519673},
  url={https://books.google.co.nz/books?id=ctM-EAAAQBAJ},
  year={2021},
  publisher={Springer Nature Singapore}
}

@article{steane1998quantum,
  title={Quantum computing},
  author={Steane, Andrew},
  journal={Reports on Progress in Physics},
  volume={61},
  number={2},
  pages={117},
  year={1998},
  publisher={IOP Publishing}
}

@article{schuld2021supervised,
  title={Supervised quantum machine learning models are kernel methods},
  author={Schuld, Maria},
  journal={arXiv preprint arXiv:2101.11020},
  year={2021}
}

@article{barthe2023gradients,
  title={Gradients and frequency profiles of quantum re-uploading models},
  author={Barthe, Alice and P{\'e}rez-Salinas, Adri{\'a}n},
  journal={arXiv preprint arXiv:2311.10822},
  year={2023}
}

@article{hubregtsen2021evaluation,
  title={Evaluation of parameterized quantum circuits: on the relation between classification accuracy, expressibility, and entangling capability},
  author={Hubregtsen, Thomas and Pichlmeier, Josef and Stecher, Patrick and Bertels, Koen},
  journal={Quantum Machine Intelligence},
  volume={3},
  pages={1--19},
  year={2021},
  publisher={Springer}
}

@article{aktar2024graph,
  title={Graph Neural Networks for Parameterized Quantum Circuits Expressibility Estimation},
  author={Aktar, Shamminuj and B{\"a}rtschi, Andreas and Oyen, Diane and Eidenbenz, Stephan and Badawy, Abdel-Hameed A},
  journal={arXiv preprint arXiv:2405.08100},
  year={2024}
}

@article{Moussa_2022,
   title={Hyperparameter Importance of Quantum Neural Networks Across Small Datasets},
   ISBN={9783031188404},
   ISSN={1611-3349},
   url={http://dx.doi.org/10.1007/978-3-031-18840-4_3},
   DOI={10.1007/978-3-031-18840-4_3},
   booktitle={Discovery Science},
   publisher={Springer Nature Switzerland},
   author={Moussa, Charles and van Rijn, Jan N. and Bäck, Thomas and Dunjko, Vedran},
   year={2022},
   pages={32–46} }

@article{bayesian_opt,
  author    = {Alaa, Ahmed M. and van der Schaar, Mihaela},
  title     = {Hyperparameter optimization for machine learning models based on Bayesian optimization},
  journal   = {ResearchGate},
  year      = {2019},
  url       = {https://www.researchgate.net/publication/332557186_Hyperparameter_optimization_for_machine_learning_models_based_on_Bayesian_optimization},
}

\end{document}